\documentclass[reprint,
]{revtex4-1}
\usepackage{amsmath}
\usepackage{amssymb}
\usepackage{physics}
\usepackage{dsfont}
\usepackage{booktabs}
\usepackage{hhline}
\usepackage{enumerate}
\usepackage{subcaption}
\usepackage{comment}
\usepackage{dirtytalk}
\usepackage{footnote}
\usepackage{hyperref}
\usepackage{graphicx}
\usepackage{amsthm}
\usepackage{makecell}
\usepackage{xcolor}
\usepackage{bold-extra} 
\renewcommand{\v}[1] 

\newcommand{\tV}{\tilde{V}}
\newcommand{\PREP}{\textrm{PREP}}
\newcommand{\SEL}{\textrm{SEL}}
\newcommand{\USP}{\textrm{USP}}
\newcommand{\SWUP}{\textrm{SWUP}}

\newcommand{\hop}{\textrm{hop}}
\newcommand{\diag}{\textrm{diag}}
\newcommand{\plog}{\textrm{polylog}}
\newcommand{\poly}{\textrm{poly}}
\newcommand{\mcalO}{\mathcal{O}}

\newcommand{\mcalN}{\mathcal{N}}
\newcommand{\mcalC}{\mathcal{C}}
\newcommand{\tb}{\tilde{b}}
\newcommand{\tc}{\tilde{c}}
\newcommand{\tH}{\tilde{H}}
\newcommand{\mD}{\mathcal{D}}
\newcommand{\ttil}{\tilde{t}}

\newcommand{\tn}{\tilde{n}}
\newcommand{\mS}{\mathcal{S}}

\newcommand{\QFT}{\text{QFT}}

\newcommand{\BE}{\text{BE}}

\newcommand{\QHT}{\textrm{QHT}}

\newcommand{\anc}{\textrm{anc}}
\newcommand{\rot}{\textrm{rot}}
\newcommand{\COMP}{\textrm{COMP}}
\newcommand{\SELECT}{\textrm{SELECT}}
\newcommand{\e}{\mathrm{e}}
\newcommand{\ii}{\mathrm{i}}
\newcommand{\hn}{\hat{n}}

\DeclareFontFamily{U}{mathx}{\hyphenchar\font45}
\DeclareFontShape{U}{mathx}{m}{n}{
      <5> <6> <7> <8> <9> <10>
      <10.95> <12> <14.4> <17.28> <20.74> <24.88>
      mathx10
      }{}
\DeclareSymbolFont{mathx}{U}{mathx}{m}{n}
\DeclareMathSymbol{\bigtimes}{1}{mathx}{"91}

\begin{document}

\title{Quantum fast-forwarding fermion-boson interactions via the polaron transform}

\author{Harriet Apel}
\author{Burak \c{S}ahino\u{g}lu}
\affiliation{PsiQuantum, 700 Hansen Way, Palo Alto, CA 94304}
\date{\today}

\begin{abstract}
Simulating interactions between fermions and bosons is central to understanding correlated phenomena, 
yet these systems are inherently difficult to treat classically.
Previous quantum algorithms for fermion-boson models exhibit computation costs that scale polynomially with the bosonic truncation parameter, $\Lambda$.
In this work we identify the efficient unitary transformation enabling fast-forwarded evolution of the fermion-boson interaction term, yielding an interaction-picture based simulation algorithm with complexity polylogarithmic in $\Lambda$.
We apply this transformation to explicitly construct an efficient quantum algorithm for the Hubbard-Holstein model and discuss its generalisation to other fermion-boson interacting models.
This approach yields an important asymptotic improvement in the dependence on the bosonic cutoff and establishes that, for certain models, fermion-boson interactions can be simulated with resources comparable to those required for purely fermionic systems.
\end{abstract}

\maketitle

\section{Introduction}

Fermions and bosons are the building blocks of physical systems, underpinning a wide range of phenomena in chemistry~\cite{conwell2000polarons, giustino2017electron, ribeiro2018polariton}, materials~\cite{bardeen1955electron, schrieffer2018theory, franchini2021polarons} and high-energy physics~\cite{schwinger1951gauge, nambu1960quasi, higgs1966spontaneous, kogut1975hamiltonian}.
To describe their coupled behaviour many model Hamiltonians ~\cite{holstein1959studies, holstein1959studies2, frohlich1954electrons, su1980soliton} have been established which capture features of entangled electron-phonon and electron-photon systems informing our understanding of composite quasiparticles such as polarons and polaritons, key to superconductivity, excitonic transport, and light-matter systems, respectively~\cite{alexandrov1992polarons, damjanovic2002excitons, ribeiro2018polariton, ghosh2020polarons,  franchini2021polarons, nosarzewski2021superconductivity}
Hence precise simulation of fermion-boson model's static and dynamic properties are crucial to better understand this emergent behavior and its manifestation in real materials and devices.

Classical simulation of fermion, boson, and fermion-boson system is notoriously challenging due to the exponential growth of the Hilbert space and the infinite dimensionality of bosonic modes.
These difficulties make such systems a natural target for quantum algorithms motivating extensive prior work~\cite{somma2002simulating, somma2003quantum, jordan2011quantum, jordan2012quantum, jordan2014quantum, somma2016quantum, somma2016trotter,  macridin2018electron, macridin2018digital, shaw2020quantum, tong2022provably}.
In any bosonic system - since they are infinite dimensional – a truncation parameter $\Lambda$ must be introduced to digitally simulate the system on a quantum computer.
In general, a reliable truncation value can be justified by energy constraints, for instance an initial state residing in a low-energy subspace imposes an upper bound on the boson occupation number due to the Hamiltonian terms~\footnote{For example, given a low-energy subspace below an energy bound $E$ and a positive-definite Hamiltonian that includes the standard bosonic energy term $\omega_0 \hat{n}_b$, this implies a boson occupation number cutoff $\Lambda = E/{\omega_0}$. To be completely rigorous, the cutoff can in fact also depend on the time evolution and specific structure of the Hamiltonian~\cite{tong2022provably}.}.
As the truncation parameter increases, so does the complexity of the quantum algorithm, both due to the purely bosonic terms and the fermion-boson interactions in the Hamiltonian.
It is therefore desirable to develop quantum algorithms that improve the scaling of the computational cost with the bosonic truncation.
At a fundamental level it is also curious to understand how including bosonic modes in a system effects the computational complexity, and how this dependence varies with the specific form of the fermion-boson interaction.

One method for reducing the cost of Hamiltonian simulation is to work in the interaction picture (IP)~\cite{low2018hamiltonian} where the Hamiltonian is decomposed as $H= H_0 + V$ where $H_0$ can be time-evolved using $o(t)$ gates (fast-forwardable) and $V$ contains the remaining terms.
The algorithm constructs a truncated and discretised Dyson series, implemented as a linear combination of products of $\mathrm{e}^{-\mathrm{i}H_0 s}$ for various times $|s| < |t|$, together with the block-encoding $\BE_{V/\alpha_V}$ of the remaining terms $V$ rescaled by a factor $\alpha_V$. 
Since the evolution under $H_0$ can be $\epsilon$-approximately simulated in $\plog(\|H_0\|s/\epsilon)$ gates, the overall gate complexity scales as ${\mathcal{O}(\alpha_V |t| \plog((\|H_0\| + \alpha_V) |t|/\epsilon))}$.
Thus, the cost is heavily determined by the rescaling factor $\alpha_V$, which depends on the portion of the Hamiltonian that remains in $V$, yielding an exponential reduction in gate complexity for diagonally dominant Hamiltonians.
Recently~\cite{tong2022provably} applied the IP-based algorithm to the Hubbard-Holstein model, together with the HHKL algorithm~\cite{haah2021quantum} that leverages geometric locality of many-body Hamiltonians to also improve the complexity scaling in the number of sites.
The authors of~\cite{tong2022provably} choose $H_0$ to be the purely bosonic part of the Hamiltonian, noting that no efficient fast-forwarding method is currently known for the fermion-boson interaction term, precluding a decomposition of $H$ that further reduces $\alpha_V$ and consequentially the complexity.

Our main technical contribution is to bridge this gap by showing that it is possible to fast-forward both the bosonic and fermion-boson interaction terms $H_b + H_{fb}$ and even extend this to include the diagonal fermionic sector $H^{\diag}_{f} + H_b + H_{fb}$.
The key ingredient for this result is an efficient unitary that diagonalises $H_b + H_{fb}$, which we identify as the displacement operator, a well-known construct in quantum optics.
By applying this operator coherently, with the displacement conditioned on the fermionic configuration, $H_{b} + H_{fb}$ is diagonalised into a non-interacting bosonic oscillator.
The construction generalises to other fermion-boson interactions where $H_{fb}$ acts diagonally on the fermionic degrees of freedom, allowing diagonalision via displacement operators $D(\alpha)$, or Bogoliubov transformations~\cite{bogolijubov, valatin},
and more generally for number-conserving quadratic bosonic Hamiltonians with Lie-algebra diagonalisation techniques~\cite{gu2021fast, kokcu2022fixed}. 
In this work we focus on the operators that are composed of displacement or controlled displacement operators.
Historically, these transformations appeared in the context of condensed-matter theory before the advent of quantum computing, with the
transform for the Holstein, Fr\"{o}hlich and single-mode SSH models appearing as Lee-Low-Pines transformation~\cite{lee1953motion}, Lang-Firsov transformation~\cite{lang1963kinetic} or the polaron transform.
Having identified this transformation, we construct efficient quantum circuits to implement the polaron transform for a range of fermion-boson interacting systems.

Leveraging this transformation, our second main result is a quantum simulation algorithm for the Hubbard-Holstein model that, up to logarithmic factors, matches the complexity of purely fermionic simulations, scaling only polylogarithmically with the bosonic truncation parameter. 
This is achieved by working in the first-quantised representation, expressing the bosonic and fermion-boson interactions terms in the $\{X,P\}$ basis rather than second-quantisation's creation and annihilation operators.
In first-quantised form, the displacement operator is particularly easy to implement, requiring only $\plog( \Lambda /\epsilon)$ gates.
When combined with recent results demonstrating efficient in $\Lambda$ evolution of the first-quantised quantum harmonic oscillator~\cite{jain2025efficientquantumhermitetransform} the cost of evolution of $H_0 = H^{\diag}_{f} + H_b + H_{fb}$ is independent of $t$ and becomes polylogarithmic in the truncation parameter, i.e., $\mcalO(\plog\Lambda)= o(\Lambda)$.
By contrast, in second quantisation, implementing the polaron transform via quantum signal processing incurs a sublinear but still algebraic dependence $\mcalO(\sqrt{\Lambda} + \log 1/\epsilon)$.
Combining this with the IP-algorithm and HHKL yields a simulating cost for the Hubbard-Holstein model of $\tilde{\mcalO}(Nt \plog(N/\epsilon)\log( \Lambda/\epsilon))$ achieving exponential improvement with respect to $\Lambda$ compared to the previous state of the art~\cite{tong2022provably}: $\tilde{O}(Nt\sqrt{\Lambda})$.
If one were to do Hamiltonian simulation by QSP~\cite{low2019hamiltonian} without HHKL the complexity of the Hubbard-Holstein model is $\tilde{O}(N^2t\Lambda)$.

\begin{table}[]
    \centering
    \begin{tabular}{|c|c|}
    \hline
        Algorithm & Complexity of Hubbard-Holstein model \\
    \hline 
    QSP~\cite{low2019hamiltonian} & $\tilde{O}(N^2t(\Lambda \omega_0 + g \sqrt{\Lambda} + \abs{U} + \abs{\mu}) + \log(1/\epsilon))$\\
    \hline
    IP + HHKL\cite{tong2022provably} & $\tilde{\mcalO}(Nt g\sqrt{\Lambda}\plog(g \Lambda/\omega_0 \epsilon, Nt\omega_0/\epsilon, \abs{U},\abs{\mu})$ \\
    \hline 
    IP + \textcolor{red}{PT} & $\tilde{\mcalO}(N^2t \log(g \Lambda/(\omega_0 \epsilon), \abs{U},\abs{\mu}))$\\
    \hline 
    IP + \textcolor{red}{PT} + HHKL & $\tilde{\mcalO}(Nt\plog(Nt\omega_0/\epsilon)\log(Ntg \Lambda/ \epsilon, \abs{U},\abs{\mu}))$ \\
    \hline
    \end{tabular}
    \caption{Comparison table of the gate complexity of simulating the Hubbard-Holstein model with $N$ sites, for time $t$ and approximation error up to $\epsilon$. 
    The different algorithms considered are quantum signal processing (QSP), and various combinations of interaction picture (IP), polaron transform (PT) and the HHKL algorithm~\cite{haah2021quantum}. $\omega_0$ is the bosonic frequency, $g$ is the fermion-boson coupling strength, $U$ is the fermionic onsite interaction, $\mu$ is the chemical potential. Note the different scalings with the bosonic truncation parameter $\Lambda$, doing QSP is linear, IP is sublinear and with the addition of PT the scaling becomes polylogarithmic -- hence the polaron transform of this work bringing exponential improvement here. }
    \label{tab:placeholder}
\end{table}

Our third result is that a similar transformation can be applied to other models such as the Dicke model, Hubbard-Fr{\"o}hlich and Su-Schrieffer-Heeger (SSH) models in electron-phonon systems.
We explicitly give the unitary transformation that leads to fast-forwarding and discuss the quantum circuit implementations and asymptotic costs.
While the fast-forwarding applies fully to the first three aforementioned models and results in an exponential improvement in the cutoff $\Lambda$, it applies to only a specific subset of the SSH models with only a single bosonic mode.

The manuscript is organized as follows.
Section~\ref{sec:PolaronTransform} studies the polaron transform in the context of the Hubbard-Holstein model, giving the explicit expressions for the unitary transformation and the resulting diagonalised Hamiltonian, both in first and second quantised form.
Section~\ref{sec:QuantumCircuit} describes the quantum circuit implementations of a full Hamiltonian simulation algorithm using state-of-the-art methods combined with the proposed fast-forwarding, and gives the asymptotic gate complexity, both in first and second quantisation.
In Section~\ref{sec:Generalizations} the polaron transform is extended to other models in quantum optics and the asymptotic costs of quantum circuit implementations are given for the fast-forwarding parts on the circuit.
We then conclude and list a few open problems in Section~\ref{sec:Conclusions}.
Finally, for the sake of completeness, the appendix consists of detailed quantum circuit implementations that can be used to unequivocally show the results in the main text parts of which may be of independent interest~\footnote{We note some (or all) of the circuits presented in the appendix may not be the most efficient way of implementing these quantum circuits when considering constant factor improvements.}.

\section{Polaron transform in the Hubbard-Holstein Model}\label{sec:PolaronTransform}

Consider a system of both fermions and bosons where the Hamiltonian has a term acting on the fermionic degrees of freedom, $H_f$, a term acting on bosonic degrees of freedom, $H_b$, and an interaction term acting between them, $H_{fb}$:
\begin{equation}
    H = H_f + H_b + H_{fb},
\end{equation}
where
\begin{multline}
H_f = -\sum_{\langle i, j \rangle, \sigma} \left(c^\dagger_{i, \sigma} c_{j, \sigma}+ c_{j, \sigma} c^\dagger_{i, \sigma} \right)
\\+ \sum_{i} \left[U\left( n_{i, \uparrow} - \frac{1}{2} \right) \left( n_{i, \downarrow} - \frac{1}{2} \right) - \mu n_i \right],
\end{multline}
is the Hubbard model involving purely electronic degrees of freedom (dofs) and 
\begin{align}\label{eq:HHinitial}
H_b & = \omega_0 \sum_{i} b^\dagger_i b_i, \\
H_{fb} & = g \sum_{i} \left(b^\dagger_i + b_i\right) \left( \hn_i - 1 \right)
\end{align}
are the bosonic and fermion-boson interaction terms, in second-quantized form where $b_i^\dagger$ ($b_i$) denotes the bosonic creation (annihilation) operator acting at site $i$, satisfying $[b_j, b^\dag_i]= \delta_{ij}$.
The fermionic ladder operators $c_{i,\sigma}^\dagger$ ($c_{i,\sigma}$) act on site $i$ and spin $\sigma\in\{\uparrow, \downarrow\}$, satisfying $\{c_{j, \sigma'}, c^\dag_{i \sigma}\}= \delta_{ij} \delta_{\sigma, \sigma'}$. $\hn_{i,\sigma} := c^\dagger_{i,\sigma}c_{i,\sigma}$ denotes the fermionic number operator and $\hn_i := \hn_{i,\uparrow} + \hn_{i,\downarrow}$.
$\omega_0$ is the bosonic oscillator frequency, $g$ is the electron-photon coupling strength, $U$ is the onsite fermionic interaction energy and $\mu$ is the chemical potential.

The Hubbard-Holstein model \cite{holstein1959studies, Ramakumar_2004} is a particularly simple Hamiltonian used to study strongly correlated electron-phonon systems. 
It captures the physics of electrons interacting with each other via on-site Coulomb repulsion as in the Hubbard model \cite{Hubbard1964}, as well as the coupling to local quantised lattice vibrations modelled by bosonic modes. 
These phonons are modelled as dispersionless as in the Holstein model \cite{holstein1959studies} and so they have the same frequency at every site.
The interaction model has no off-site coupling between electrons and phonons -- a key simplification over more complex models such as Hubbard-Fr\"{o}hlich and SSH models (discussed in Sections \ref{subsec:Frohlich} and \ref{subsec:SSH}).

In any type of bosonic system -- since bosons are infinite dimensional -- a truncation must be introduced to digitally simulate the system on a quantum computer. 
We explicitly consider both first and second quantisation formulations of bosonic system, however the remainder of the paper we will focus more on first quantisation as this gives the best Hamiltonian simulation complexity -- initially the Hamiltonian was presented in second quantisation for familiarity.
In first quantisation the bosonic modes are described in terms of position and momentum operators, and both take $M$ discrete values. 
Second quantisation describes bosonic modes in terms of occupation numbers, and $\Lambda$ denotes the maximum occupation number.
The truncation parameters are given different symbols as they are not equivalent although we will see later that they have the same scaling $\Lambda \in \Theta(M)$.

The choice of quantisation will impact which terms of the Hamiltonian can be efficiently implemented. 
However, in both formulations the boson-fermion interaction term is not naturally diagonal and so contributes to the norm of $V$ when implementing Hamiltonian simulation using the interaction picture algorithm. 
The unitary that implements the polaron transform is used to transform the boson-fermion interactions in both models so that it is fast-forwardable:
\begin{equation}
    \mathcal{D} := \bigotimes_i \left[\sum_{n_i=0}^2 D(\alpha_{n_i}) \otimes \ketbra{n_i}{n_i} \right],
\end{equation}
where $D$ is the displacement operator acting on the bosonic degree of freedom at site $i$.
It is parametrised by $\alpha_{n_i}$ which quantifies the displacement and depends on the number of fermions $n_i$ at the same site. 
A tensor product of these displacement operators is applied at every site $i \in \{1,2,\dots,N \}$.
We consider a spin $1/2$ model so each $n_i$ can take values from $\{0,1,2\}$.
This operator, $\mD,$ is also called the the Lang-Firsov transformation in Ref.~\cite{lang1963kinetic, hohenadler2007lang}.

\subsection{Transformation in first quantisation}\label{subsec:1QPolaronTransform}

The boson and fermion-boson Hamiltonian terms of the Hubbard-Holstein model in first quantisation are given by,
\begin{align}\label{eq:HHinBosons1stQ}
H_b &= \frac{\omega_0}{2} \sum_i X^2_i + P^2_i,\\
H_{fb} &= g\sqrt{2}\sum_i X_i (\hat{n}_i - 1),
\end{align}
where $X_i$ and $P_i$ are position and momentum operators of the bosonic oscillator placed on site $i$, respectively, with the canonical commutation relations $[X_i,P_{j}]= \ii \delta_{i,j}$. 
Note that the bosonic frequency could be site-dependent, i.e., $H_b= \sum_i \frac{\omega_i}{2} (X^2_i + P^2_i)$ , which would not change the overall complexity, but for simplicity we consider uniform frequency here.
The polaron transform in this representation is given by
\begin{align}\label{eq:DisplacementOperator1stQ}
D(\alpha)= e^{-i\alpha P}
\end{align}
which results in a displacement in the position of the oscillator, i.e., 
\begin{align}\label{eq:ActionOfDisplacement1stQ}
D(\alpha)^\dag X D(\alpha)= X + \alpha \mathds{1} := \tilde{X}.
\end{align}
The momentum operator is invariant under the displacing operator for $\alpha$ real:
\begin{equation}
    \forall \alpha \in \mathbb{R} \: :D(\alpha)^\dag P D(\alpha)= P.
\end{equation}

We can write the sum $H_0 = H_b + H_{bf}$ in terms of the shifted operators $\tilde{X}$ and $\tilde{P}$:
\begin{subequations}
\begin{align}
H_b + H_{fb}= \frac{\omega_0}{2} \sum_{i}&\Bigg[X_i^2 + P_i^2 + \frac{2\sqrt{2}g}{\omega_0} X_i \otimes (\hat{n}_i - 1) \Bigg] \\
= \frac{\omega_0}{2} \sum_{i} \sum_{n_i}&\Big[\tilde{X}_{i, n_i}^2 + \tilde{P}_{i,n_i}^2 - \abs{\alpha_{n_i}}^2 \mathds{1}\Big]\otimes \ketbra{n_i}_i.
\end{align}
\end{subequations}
The displacement required depends on the fermionic occupation:
\begin{align}
\alpha_{n_i} = \begin{cases}
+\sqrt{2}g/\omega_0,\; &\textrm{if} \; n_i=2,\\
0 ,\; &\textrm{if} \; n_i = 1,\\
-\sqrt{2}g/\omega_0,\; &\textrm{if} \; n_i=0.
\end{cases}
\end{align}

The sum can then be written explicitly in terms of the polaron transform:
\begin{widetext}
\begin{align}
H_b + H_{fb} &= \underbrace{ \mathcal{D}^\dagger  \left( \frac{\omega_0}{2} \left[{X}_i^2 + {P}_i^2 \right] \otimes \mathds{1} \right) \mathcal{D}}_{H'_{b}} 
+ \underbrace{\left(- \frac{\omega_0}{2} \sum_i \sum_{n_i=0}^2 |\alpha_{n_i}|^2 \mathds{1} \otimes \ketbra{n_i}_i \right)}_{H'^{\text{diag}}_{f}}\\
&= \mD^\dag (H_b + H'^{\diag}_f) \mD.
\end{align}
\end{widetext}
Hence the fermion-boson interaction term summed with the bosonic term is transformed by the polaron transform into a sum of purely bosonic diagonal Hamiltonian and a purely fermionic diagonal term $H'^{\text{diag}}_{f}$.

\subsection{Transformation in second quantisation}\label{subsec:2QPolaronTransform}

The boson and fermion-boson Hamiltonian terms of the Hubbard-Holstein model in second quantisation are given by,
\begin{align}\label{eq:HHinBosons2Q}
H_b & = \omega_0 \sum_{i} b^\dagger_i b_i, \\
H_{fb} & = g \sum_{i} \left(b^\dagger_i + b_i\right) \left( \hn_i - 1 \right).
\end{align}
The polaron transform in this representation is
\begin{align}\label{eq:DisplacementOperator2Q}
D(\alpha)= \exp(\alpha b^\dagger - \alpha^* b),
\end{align}
which results in shifted bosonic creation/annihilation operators, i.e., 
\begin{align}\label{eq:HHDisplacedAnnihilation}
\tb &= D(\alpha)^\dag b D(\alpha)= b + \alpha \mathds{1}\\
\tb^\dagger &= D(\alpha)^\dag b^\dagger D(\alpha)= b^\dagger + \alpha^\star \mathds{1}
\end{align}

Again we start by expressing the sum of the boson and fermion-boson Hamiltonian components in terms of the shifted operators,
\begin{subequations}
\begin{align}
H_b + H_{fb} = \omega_0 \sum_{i}&\Big[ b^\dag_i b_i + \frac{g}{\omega_0} (b^\dag_i + b_i) (n_i - 1) \Big]\\
\nonumber = \omega_0 \sum_{i}&\Big[ b^\dag_i b_i + \frac{g}{\omega_0} (b^\dag_i + b_i) \otimes \ketbra{11}_i\\
&- \frac{g}{\omega_0} (b^\dag_i + b_i) \otimes \ketbra{00}_i \Big]\\
=\omega_0 \sum_i & \sum^2_{n_i=0} (\tb^{\dag}_{i, n_i} \tb_{i, n_i} - |\alpha_{n_i}|^2 \mathds{1}_i ) \otimes \ketbra{n_i}_i.
\end{align}
\end{subequations}
The displacement required for the boson at site $i$, again depends on the fermionic occupation number at the same site, $n_i$.
The required displacements are,
\begin{align}\label{eq:HHDisplacedAlpha}
\alpha_{n_i} = \begin{cases}
+g/\omega_0,\; &\textrm{if} \; n_i=2,\\
0 ,\; &\textrm{if} \; n_i = 1,\\
-g/\omega_0,\; &\textrm{if} \; n_i=0.
\end{cases}
\end{align}

The sum can again be expressed using the polaron transform as a bosonic Hamiltonian plus a purely fermionic diagonal term:
\begin{widetext}
\begin{subequations}
\begin{align}
H_b + H_{fb} &= \underbrace{\mathcal{D}^\dagger \left(\omega_0 \sum_i b^{\dag}_{i} b_{i} \otimes \mathds{1} \right)  \mathcal{D}}_{H'_{b}}
+ \underbrace{\left(- \omega_0 \sum_i \sum_{n_i=0}^2 |\alpha_{n_i}|^2 \mathds{1} \otimes \ketbra{n_i}_i \right)}_{H'^{\text{diag}}_{f}}\\
&= \mD^\dag (H_b + H'^{\text{diag}}_f) \mD.
\end{align}
\end{subequations}
\end{widetext}
Hence the fermion-boson interaction term summed with the bosonic term is again transformed by the polaron transform into a sum of purely bosonic diagonal Hamiltonian and a purely fermionic diagonal term $H'^{\text{diag}}_{f}$.

\subsection{Fast-forwarding in the Hubbard-Holstein Model}

The fermionic term of the Hubbard-Holstein model consists of a hopping term, $H^\text{hop}_f$, that must be block-encoded and a diagonal term, $H^\text{diag}_f$, that is trivially fast-forwardable:
\begin{align}
H_f =& \underbrace{-\sum_{\langle i, j \rangle, \sigma} \left(c^\dagger_{i, \sigma} c_{j, \sigma} + c_{j, \sigma} c^\dagger_{i, \sigma} \right)}_{H^\text{hop}_f}\notag \\
&\underbrace{+ \sum_{i} \left[U\left( n_{i, \uparrow} - \frac{1}{2} \right) \left( n_{i, \downarrow} - \frac{1}{2} \right) - \mu n_i \right]}_{H^\text{diag}_f}.
\end{align}
By noting that $[\mD, H'^\diag_f]= 0$ and similarly $[\mD, H^\diag_f]= 0$, it follows that:
\begin{subequations}\label{eqn:form_of_h0}
\begin{align}
H_b + H_{fb} + H^{\diag}_f&= H'_{b} + H'^{\diag}_f + H^{\diag}_f\\
& =  \mathcal{D}^\dag ( H_{b} + H'^{\diag}_f + H^{\diag}_f) \mathcal{D}\\
& = \mathcal{D}^\dag \tH_0 \mathcal{D}.
\end{align}
\end{subequations}
Hence we claim the bosonic, fermion-boson interaction and diagonal fermionic terms can all be fast-forwarded in the interaction picture.
The split for the Hamiltonian is then:
\begin{equation}\label{eqn:hamiltoniansplit}
    H = \mathcal{D}^\dag \tH_0 \mathcal{D} + V
\end{equation}
where $V = H_f^\text{hop}$.

One way of using the polaron transform for simulation is to use the two terms in Eq.~\eqref{eqn:hamiltoniansplit} as inputs to the interaction picture algorithm.
This would involve repeated applications of the transform to transition into and out of the basis that diagonalises the fermion–boson interaction.
However, since in between the non-diagonalisable fermionic interaction will be block-encoded it is favourable to just apply this transformation once over the whole Hamiltonian.
\begin{align}
\tH = \mD H \mD^\dag= \tilde{V} + \tilde{H}_0
\end{align}
where we now block-encode directly the transformed operator,
\begin{equation}
    \label{eq:1QtildeV}\tilde{V}= - \sum_{\langle i,j \rangle, \sigma} c^{\dag}_{i, \sigma} c_{j, \sigma} e^{-i g\sqrt{2}(P_i - P_j)/\omega_0}.
\end{equation}
We can then modify our quantum algorithm so that we first apply the unitary $\mD$, and then perform Hamiltonian simulation with $\tH$, and finally apply $\mD^\dag$.
Remark that the new interaction term $\tV$ can be block-encoded with almost the same cost as of the block-encoding $V$, due to the diagonal fastforwardable terms $e^{-\ii g\sqrt{2} P_i/\omega_0}$ and $e^{+\ii g\sqrt{2} P_j/\omega_0}$ on site $i$ and $j$, respectively.
This is straightforward in the first-quantized form.
Hence, it is then left to demonstrate that $\tH_0$ and the transform $\mathcal{D}$ have efficient circuit implementations.

\section{Quantum circuit implementations and asymptotic complexity}\label{sec:QuantumCircuit}

\begin{figure*}[hbt!]
    \centering
    \includegraphics[width=0.9\linewidth]{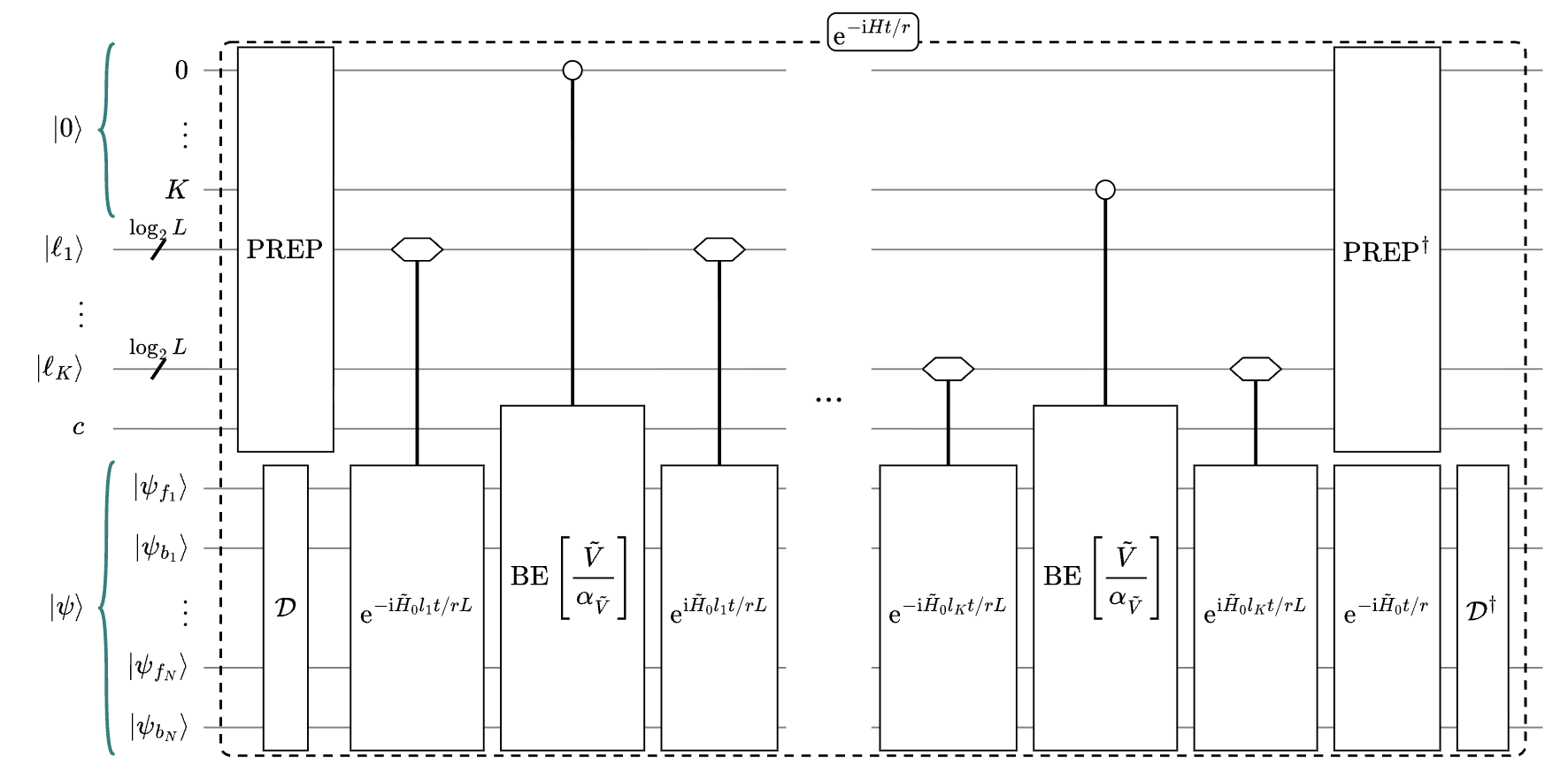}
    \caption{High-level circuit diagram for the interaction picture algorithm for quantum simulation of $\mathrm{e}^{-\mathrm{i}Ht}$ where $H$ is the Hubbard-Holstein model, the piece denoted is $\mathrm{e}^{-\mathrm{i}H_0t/r} \mathcal{T} \left[\mathrm{e}^{-\mathrm{i}\int_{0}^{t/r} V(s) ds}\right]$, which must be repeated $r$ times to evolve by time $t$.
    $\bigotimes_{i=1}^N\ket{\psi_{f_i}}\otimes\ket{\psi_{b_i}}$ denotes the $N$ site system register where each site has a single qubit to represent the fermionic degree of freedom and a $\lceil \log_2(\Lambda) \rceil$ qubit register to hold the bosonic degree of freedom truncated at $\Lambda$.
    $\tV= \mD V \mD^\dag$ and $\tilde{H}_0= \mD H_0 \mD^\dag$ denote the polaron transformed Hamiltonian terms given in Eq.~\eqref{eq:1QtildeV} and Eq.~\eqref{eqn:form_of_h0}, respectively.
    The time-ordered exponential is approximated by the Dyson expansion truncated at commutators of depth $K$ requiring a $K$-qubit ancilla register to label the integrals in the expansion, each ancilla qubit requires a $\log_2 L$ ancilla register to facilitate the approximation of the integral by a $L$-term sum. 
    Using the polaron transform $\mathcal{D}$ we can efficiently implement evolution under the bosonic, diagonal fermionic and the bosonic-fermionic interaction directly, denoted $\tilde{H}_0$ whereas the evolution under the rest of the fermionic term is block encoded requiring an ancillary register labeled $\ket{c}$. 
    See \cite{kan2025optimized} for more detailed implementation of the PREP and control structure to implement this truncated Dyson series expansion.}
    \label{fig:algo}
\end{figure*}

We apply the quantum simulation using the interaction picture algorithm, first given in Ref.~\cite{low2018hamiltonian}.
Specifically time evolving under $H$ by splitting into $H = H_0 + V$ so that:
\begin{equation}
    \mathrm{e}^{-\mathrm{i}Ht} = \mathrm{e}^{-\mathrm{i}H_0t} \mathcal{T} \left[e^{-i\int_0^t V(s) ds}\right]
\end{equation}
where we now need a time-ordered exponential of the now time-dependent perturbation operator:
\begin{equation}\label{eqn:vs}
    V(s) := \mathrm{e}^{\mathrm{i}H_0s}V\mathrm{e}^{-\mathrm{i}H_0s}.
\end{equation}
The total simulation time is divided into $r$ equal segments and additive errors then require each simulation by time $t/r$ to be accurate up to $\epsilon' = \epsilon/r$:
\begin{equation}
    \mathrm{e}^{-\mathrm{i}Ht} \approx \prod_{\ell = 0}^{r-1}\mathrm{e}^{-\mathrm{i}H_0t/r} \mathcal{T} \left[\mathrm{e}^{-\mathrm{i}\int_{0}^{t/r} V(s) ds}\right].
\end{equation}
The time-ordered exponential of equation \eqref{eqn:vs} can be implemented following \cite{kieferova2019simulating} by a $K$-truncated Dyson expansion where each of the $K$ integrals is approximated by a sum of $L$ terms.
To bound this approximation by $\epsilon'$ the truncation parameters scale as 
\begin{align}
    K &\in \Theta \left( \frac{\log(1/\epsilon')}{\log\log(1/\epsilon')} \right),\\
    L &\in \Theta \left(\frac{t^2 \max_{s\in\{0,t\}}\norm{dV(s)/dt}}{\epsilon' r^2} \right).
\end{align}
This also requires $r\geq \norm{V(s)}t$ and so we take $r = \norm{V}t/t_0$, where $t_0$ is a constant such that $\sum_{k=0}^K t_0^k/k! =2$ and so can be taken to be $\approx0.7$ in practice.
The truncated Dyson expansion is then implemented by a linear combination of unitaries (LCU) \cite{berry2015simulating} where $\tilde{V} = \mathcal{D}^\dagger H_f^\text{hop}\mathcal{D}$ is block-encoded to a unitary (circuit detailed in Appendix \ref{app:BEHfhop}), and $\mathrm{e}^{-\mathrm{i}H_0 s}$ are applied for different time parameters $s$ controlled with an ancilla register.
The high level circuit for this algorithm is given in figure \ref{fig:algo}.

The two key subroutines black-boxed in this algorithm are the polaron transform $\mathcal{D}$ and the evolution under the fast-forwardable Hamiltonian $\tilde{H}_0$.
The circuits for these routines are described for both bosonic representations in the following sections.
Letting $C(\cdot)$ denote the cost of a given subroutine, the asymptotic scaling for the simulation algorithm described above is:
\begin{multline}
    \mcalO\Bigg(r\times \Big(\log(r/\epsilon) \Big[C(\textup{BE}(\mD^\dagger V \mD/\alpha_V)) + C(\mathrm{e}^{-\mathrm{i}\tilde{H_0}t/r})\\ 
   + \log(t/\epsilon) \Big]\Big) + C(\mD)\Bigg)
\end{multline}
Implemented as given, the scaling with system size $N$ would be quadratic since $C(\textup{BE}(\tV/\alpha_{\tV}))\in \mcalO(N)$ (we will see in the next section that naively $C(\mathrm{e}^{-\mathrm{i}\tilde{H_0}t/r})$ also has this scaling) and $r\in \mcalO(\norm{V}t)\in \mcalO(N)$.
However, we can achieve the same $\mcalO(N\log(N))$ scaling with system size by following \cite[Appendix E]{tong2022provably} in employing the Haah-Hastings-Kothari-Low (HHKL) decomposition \cite{haah2021quantum}.

The HHKL decomposition reduces the complexity of simulating a geometrically local Hamiltonian by approximating it as a product of short-time evolutions on overlapping portions of the system. 
The original result \cite[Lemma 6]{haah2021quantum} states that local terms should be bounded by a constant whereas the bosonic terms in the Hubbard-Holstein model scale with the truncation parameter. 
However, in the same way as \cite{tong2022provably}, these are `on-site' interactions and so do neither impact the Lie-Robinson velocity of the system \cite[Lemma 13]{tong2022provably} nor disrupt application of the HHKL result~\footnote{Note that in systems where there are fermion-boson or boson-boson interactions \emph{between} sites in the lattice, this application does not necessarily extend.}.
Therefore, we consider simulation blocks of size $\mathcal{O}(\textup{polylog}(Nt\epsilon^{-1}))$ for time $\tau\in \mathcal{O}(1)$ up to error $\mathcal{O}(N^{-1}t^{-1}\epsilon)$.
The complexity of simulating such a block using the interaction picture algorithm described in Figure \ref{fig:algo} is given by
\begin{multline}
    \mcalO\left(\plog(Nt\omega_0/\epsilon) \left[C(\textup{BE}(\tV/\alpha_{\tilde{V}})) + C(\mathrm{e}^{-\mathrm{i}\tilde{H_0}\tau})\right]\right.\\
    \left. + C(\mathcal{D})\right)
\end{multline}
where now $\mathcal{D}$, $\mathrm{e}^{-\mathrm{i}\tilde{H_0}\tau}$ and $\textup{BE}(\tilde{V}/\alpha_{\tilde{V}})$ act on systems of size $\mathcal{O}(\textup{polylog}(Nt\omega_0\epsilon^{-1}))$
and we require $\mathcal{O}(Nt\omega_0)$ such simulations.
In the following sections we complete this complexity analysis by providing big-O analysis of the polaron transform, block encoding and fast-forwardable evolution in first and second quantisation.
The details of the circuit construction for these various pieces is postponed to the appendix.

\subsection{Circuits in first quantisation}

To represent first-quantised operators on a digital quantum computer we must first chose how the discretised position ($X^d$) and momentum ($P^d$) operator are defined.
We follow~\cite{somma2016quantum} -- also adopted in \cite{jain2025efficientquantumhermitetransform} -- where the bosonic Hilbert space on a given site has dimension $M$ (taken to be even for simplicity) and the discretised position operator is given by the following diagonal matrix,
\begin{equation}
    X^d : = \sqrt{\frac{2\pi}{M}}\frac{1}{2}\begin{pmatrix}
    -M & 0 & \dots 0 \\
    0 & (-M +2) &\dots & 0\\
    \vdots & \vdots &\ddots & \vdots\\
    0 & 0 &\dots & (M-2)
    \end{pmatrix}.
\end{equation}
The discretised momentum operator is defined taken as the `centered discrete Fourier transform' of $X^d$:
\begin{equation}\label{eqn:centeredQFT}
    P^d:=\left(F_c^d \right)^{-1} X^d \left(F_c^d \right),
\end{equation}
where $\left[F_c^d \right]_{k,l} := \frac{1}{\sqrt{M}} \exp \left(\mathrm{i} 2\pi kl/M \right)$ is a modified discrete Fourier transform where the frequency indices are now arranged symmetrically about zero, $k,l \in \{-M/2,\dots,M/2-1 \}$.

This truncation is physically meaningful in that up to an energy cutoff the quantum harmonic oscillator eigenvectors are well approximated.
Whereas the eigenfunctions of the continuous quantum harmonic oscillator are expressed in terms of the $m$-th physicists' Hermite polynomials $H_m(x)$:
\begin{equation}
  \psi_m(x) := \frac{1}{\sqrt{m!2^m\sqrt{\pi}}}\mathrm{e}^{-x^2/2}H_m(x).  
\end{equation}
\cite{somma2016quantum} shows that with the above discretisation
\begin{equation}\label{eqn: QHO eigenstates}
    \ket{\psi_m^d}:= \left(\frac{2\pi}{M} \right)^{1/4} \sum_{x=-M/2}^{M/2-1} \psi_m(x)\ket{x}
\end{equation}
are $\exp(-\Omega(M))$-approximate eigenfunctions of the discrete quantum harmonic oscillator with eigenvalues close to those expected in the continuous case, i.e.
\begin{equation}
    \norm{H^d_b \ket{\psi_m^d} - (m+1/2)\ket{\psi_m^d}}^2 = \exp(-\Omega(M)).
\end{equation}

The above holds for all $m\leq cM$ for some constant fraction of the energy spectrum $c\in(0,1)$.
The analytic results require $c<\pi/6$ so only provably hold in the low-energy spectrum \cite[Appendix A]{somma2016quantum} but numerics find that in fact the eigenfunctions are still very good approximations up to $c=3/4$ \cite[Figure 1,2]{somma2016quantum}.
However, for scaling argument it follows that the energy cutoff $\Lambda \in \Theta (M)$ and this can be used to compare the circuit complexities in first and second quantisations. In the following we drop the superscript $d$ to denote discretisation and it is implicit that  all first quantised operators and states are discretised up to cutoff $M$.

The qubits in first quantised circuits are encoded in the $X$ basis.
The single site displacement operator is $D(\alpha) = \mathrm{e}^{-\mathrm{i}\alpha P}$.
A $\mathrm{QFT}$ efficiently transforms from position to momentum basis and then the displacement is efficient to implement as an exponential of a diagonal operator (Appendix \ref{appsubsec:FFDiagonalHamiltonians}).
The full transform is slightly more complex as the magnitude of the displacement depends on the fermionic occupation. 
The number of T gates scales linearly in the number of sites $N$ (or when applying HHKL the size of the block) but only logarithmically in the bosonic cutoff $M$.
Hence, in first quantisation, for a single HHKL block of size $\plog(Nt/\epsilon)$, we have a circuit complexity ${\mathbf{C_{1Q}}(\mathcal{D})\in \mathcal{O}(\plog(Nt\omega_0/\epsilon)\log(M))}$.
A detailed circuit implementation for the block of size $N$ is given in Appendix \ref{appsubsec:DS1Q}.

Recall from Eq. \ref{eqn:form_of_h0}: $\tilde{H_0} = H_b+\tilde{H}_f^{\text{diag}}$ where ${H_b = \frac{\omega_0}{2}\sum_i X_i^2 + P_i^2}$ and $\tilde{H}_f^{\text{diag}} = \sum_i U(n_{i,\uparrow}-\frac{1}{2})(n_{i,\downarrow}-\frac{1}{2}) - \mu n_i - \frac{\omega_0}{2}\sum_{i}\sum_{n_i=0}^{2}\abs{\alpha_{n_i}}^2 \ketbra{n_i}$.
Since $H_b$ and $\tilde{H}_f^{\text{diag}}$ act non-trivially on disjoint degrees of freedom they commute:
\begin{equation}
    \mathrm{e}^{-\mathrm{i}\tilde{H}_0 \tau} = \mathrm{e}^{-\mathrm{i} H_b \tau}\mathrm{e}^{-\mathrm{i}\tilde{H}_f^{\text{diag}} \tau}
\end{equation}
and they can be applied in parallel if desired.
$\mathrm{e}^{-\mathrm{i}\tilde{H}_f^{\text{diag}} \tau}$ is diagonal and so its complexity scales linearly with the block size -- $\mathcal{O}(\textrm{polylog}(Nt\omega_0/\epsilon,\abs{U},\abs{\mu},g^2/\omega_0))$ for a single HHKL block. 

For $\mathrm{e}^{-\mathrm{i}H_b \tau}$,  we use the recent result of \cite[Theorem 5]{jain2025efficientquantumhermitetransform} to factorise this evolution into applications of $\mathrm{e}^{-\mathrm{i} a(\tau \omega_0/2) X_i^2}$ and $\mathrm{e}^{-\mathrm{i} b(\tau \omega_0/2) P_i^2}$ where one mildly increases the cutoff to $M\in \Theta(\Lambda \log(\Lambda))$, and $a(\cdot)$ and $b(\cdot)$ are trigonometric functions~\footnote{The prior state of the art was based on the Trotter-Suzuki formula and required sub-polynomial but super-polylogarithmic complexity scaling in the bosonic cutoff \cite{somma2016quantum} (Appendix \ref{append:prior1Q}).}. 
These unitaries both have efficient implementations: again scaling linearly in the number of sites but only logarithmically in the bosonic cutoff.
The implementation requires only a constant number ($3$) of calls to such unitaries to exponentially well approximate the full evolution~\footnote{The constant of $3$ is for the case where $\tau\leq \pi/2\omega_0$, this is increased to $5$ in the case where this inequality is violated, and some quantum arithmetic is needed.} and hence ${\mathbf{C_{1Q}}(\mathrm{e}^{-\mathrm{i}\tilde{H}_0\tau})\in \mcalO(\plog(Nt\omega_0/\epsilon)\log(M,\abs{U},\abs{\mu}))}$.
Relevant details of this as well as circuit descriptions of $\mathrm{exp}(-\mathrm{i}(X^2 s)$, $\mathrm{exp}(-\mathrm{i}(P^2 s)$ and $\mathrm{exp}(-\mathrm{i}{\tilde{H}}_f^{\diag})$ are given in Appendix \ref{appen:H01Q}.

Finally we consider the cost of the remaining part of the Hamiltonian that is not fast-forwardable: the block encoding of $\tilde{V} = \mathcal{D}^\dagger V \mathcal{D}$.
The circuit and costing for block encoding $V$ of the full lattice with rescaling factor $\alpha_{\tV} = 4N$ is given in detail in Appendix \ref{app:BEHfhop}.
The Toffoli cost for this is $\mathcal{O}(N + \log_2(N/\epsilon))$, which when considering the cost of one HHKL block becomes $\plog(Nt\omega_0/\epsilon)$ -- with the rescaling factor becoming $\alpha_V^{\text{HHKL}} \in O(\plog(Nt\omega_0/\epsilon))$.
Block-encoding the transformed hopping Hamiltonian, $\tilde{V}= - \sum_{\langle i,j \rangle, \sigma} c^{\dag}_{i, \sigma} c_{j, \sigma} e^{-i g\sqrt{2}(P_i - P_j)/\omega_0}$ only introduces minor modifications to the circuit and an additive $\mcalO(\log(g M/(\omega_0 \epsilon')))$ terms to the complexity per block-encoding of $\epsilon'$ precision.
Hence, the final cost of block-encoding a HHKL block in first quantisation is ${\mathbf{C_{1Q}}(\textup{BE}(\tilde{V}/\alpha_{\tilde{V}}))\in \mcalO(\plog(Nt\omega_0/\epsilon) + \log(g M/(\omega_0 \epsilon')))}$.
Note that the effect of absorbing the transform into the block-encoding rather than applying directly is an additive $\log(M)$ gates due to the swap-up circuit structure -- however when substituted back into the overall expression this doesn't change the scaling due to other contributors.

Hence we arrive at an implementation for simulating the Hubbard-Holstein with only
\[\tilde{\mcalO}(Nt \plog(Nt\omega_0/\epsilon)\log(g M Nt/(\omega_0 \epsilon),\abs{U},\abs{\mu}))\]
operations.

\subsection{Quantum Hermite transform and circuits in second quantisation}

We have presented the polaron transform to simplify the fermion-boson interaction for the Hubbard-Holstein model in both first and second quantisations, in Section~\ref{subsec:2QPolaronTransform}.
While the second-quantised formulation provides an alternative formulation of the model, it results in no asymptotic advantage/disadvantage in complexity compared with the first-quantised treatment, owing to the availability of the efficient quantum Hermite transform (QHT)~\cite{jain2025efficientquantumhermitetransform}.
Since different simulation strategies and challenges are present in each framework, we include a high-level description of the circuits for the Hubbard-Holstein in Appendix~\ref{app:BlockEncodingGeneral}, as for different models, a combination may be most advantageous. 

The truncation in second quantisation is straightforward with the qubits encoding the bosonic occupation number $\ket{m}$ with a bosonic register of $\log_2(\Lambda)$ qubits for each site able to represent occupation numbers up to $\Lambda$.
The single-site displacement operator, $D(\alpha)= \exp(\alpha b^\dagger_i - \alpha^* b_i)$ is not diagonal and so a direct implementation is more complex than in the first-quantised case, unless one uses a quantum Hermite transform, which is a unitary that performs the mapping
\begin{align}
\ket{m} \mapsto \ket{\psi_m}= \left(\frac{2\pi}{M} \right)^{1/4}\sum^{M/2-1}_{x=-M/2} \psi_{m}(x) \ket{x}
\end{align}
for all $n \in [1, \Lambda]$.
Recent work ~\cite{jain2025efficientquantumhermitetransform} proves that a quantum circuit consisting of $\plog(\Lambda)$ gates can perform this transformation for $\Lambda= M/\ln M$.
Utilising this transformation, the cost of implementing $\mD$ then becomes equivalent to the cost of implementing $\mD$ up to an additive $\mcalO(N)$ many QHTs.
Alternatively, another circuit implementation up to error $\epsilon$ can be achieved using Hamiltonian simulation by qubitisation \cite{low2019hamiltonian, Berry_2024} of evolution operator $D(\alpha) = \mathrm{e}^{-\mathrm{i}Kt}$ where the iterate of this operator $W(K)$ is constructed by the usual linear combination of unitaries.
The single-site displacement operator is constructed using Hamiltonian simulation via QSP and applied in parallel across all $N$ sites so the final complexity scaling is $\mathcal{O}(N(\frac{g\sqrt{\Lambda}}{\omega_0}) + \log(\frac{1}{\epsilon}))$.
Circuits for both $W(K)$ and $\mathcal{D}$ are given in Appendix \ref{appsubsec:DS2Q}.

The evolution under the transformed free Hamiltonian $\tilde{H}_0$ is immediate in second quantisation because both the fermionic and bosonic terms are diagonal in the occupation basis, allowing straightforward phase rotations on each register. 
The gate complexity for this diagonal evolution scales as $\mathcal{O}(N\log(1/\epsilon))$.

\subsection{Remarks on the complexity of the polaron transform and the overall algorithm}

The polaron transformation plays a crucial role in transforming the Hamiltonian so that the fermion-boson interaction terms are \emph{eliminated}. 
So, it is important to understand the complexity of this subroutine and its place in the full quantum algorithm.

First, we remark that the polaron transform need not be performed on the quantum computer.
One could directly implement a Hamiltonian simulation algorithm using the transformed Hamiltonian $\tH$ as well.
This would however require preparing the initial state in the transformed basis, as well as modifying the observables and the final state to be in the transformed basis, in case the final goal is to design an observable or overlap estimation algorithm.
For example, in an interaction-picture-based algorithm, this corresponds to pushing the unitaries, $\mD$ and $\mD^\dag$ at the beginning and end of the quantum circuit given in Fig.~\ref{fig:algo}, to the observables and initial/final states:
\begin{align}
        \bra{\psi_0} \mathrm{e}^{\mathrm{i}Ht} O \mathrm{e}^{-\mathrm{i}Ht} \ket{\psi_0} = \bra{\psi_0} \mD \mathrm{e}^{\mathrm{i}\tH t} \tilde{O} \mathrm{e}^{-\mathrm{i}\tH t} \mD^\dag \ket{\psi_0}
    \end{align}
where $\tilde{O}= \mD^\dag O \mD$.
However, in adiabatic Hamiltonian simulation or if the initial state is not classically describable, e.g., the result of some other Hamiltonian simulation, it is essential to implement the transform on the quantum computer as we have described. 

Second, the polaron transform is a tensor product over different sites $i$, hence is embarrassingly parallelisable over the sites~\cite{wikipedia:embarrassingly_parallel}.
Hence if the displacement operator per site can be implemented classically ($\Lambda$ is not too large), it is just a matter of memory to implement the full polaron transform.
For example, if the quantum state can be described by a low enough bond-dimension MPS over different sites, then the polaron transform only leads to updating the local tensor at each site.

Third, the formalism (e.g., first or second-quantised) we work in can result in a difference in the complexity of the polaron transform.
The polaron transform at each site is a direct sum of diagonal unitaries in the first-quantised formalism, and even further, it is the simple operator $\mathrm{e}^{-\mathrm{i}P\alpha_i}$ where $\alpha_i$ depends on the fermionic occupation on site $i$.
Hence it is again embarrassingly parellelisable over the qubits that describe the bosonic dof.
However, this requires one to be in the momentum basis, and within the overall algorithm this implies application of QFT, which is exponentially faster on a quantum computer compared to its classical counterpart FFT.
Similarly, purely in the second-quantised formalism, the polaron transform is no longer a simple operator, and would take $\poly(\Lambda)$ to implement classically.
However, as we have seen, on a quantum computer, one could make use of the efficient QHT to implement it using only $\plog(\Lambda)$ gates.
Hence, while the polaron transform itself can be implemented efficiently classically in the right basis and on special states (which can be compressed to fit in classical memory), making use of it in the overall Hamiltonian simulation algorithm in general requires it to be implemented on the quantum computer for the improvements shown in the manuscript to hold.

\section{Fast-forwarding in other models}\label{sec:Generalizations}

In this section, we show that the polaron transform can be used in other models to fast-forward the fermion-boson interaction term.
We first examine a generalisation of the Jaynes-Cummings~\cite{jaynes2005comparison} model, i.e., the Dicke model~\cite{dicke1954coherence}.
For a recent comprehensive review on these models, see Ref.~\cite{larson2021jaynes}.
We then show the fast-forwarding of the interactions of the Hubbard-Fr\"{o}hlich model~\cite{frohlich1954electrons}, which captures the behaviour of `large polarons' with a geometrically nonlocal interaction, and can be seen as a generalised version of the Holstein model.
Finally, we examine the Su-Schrieffer-Heeger (SSH) model~\cite{su1980soliton}, which captures electron-polaron interactions by coupling the phonon annihilation/creation to electron hopping.
We remark that the polaron transform can be utilised for fast-forwarding in the SSH model when a single bosonic mode is present, yet it is unclear whether the same or a more general transform will provide fast-forwarding for the general case consisting of multiple bosonic modes.

\subsection{Dicke model}\label{subsec:Dicke}

The Dicke Hamiltonian models the collective interaction between a single quantised bosonic mode and a system of $N$ two-level systems (atoms)~\cite{dicke1954coherence}.
Thus it is especially relevant when exploring phenomena in quantum optics and cavity QED.
The Hamiltonian reads as follows~\cite{kirton2019introduction}:
\begin{align}\label{eq:HamiltonianDicke}
H_D= \omega_0 b^\dag b \otimes \mathds{1} + \Omega \mathds{1} \otimes S_z + g (b + b^\dag) \otimes S_x.
\end{align}
Here $b$, $b^\dag$ are the creation annihilation operators for the bosonic mode in second quantisation with frequency $\omega_0$.
The collective spin operators, $S_z= \sum^N_{j=1} Z_j$ and $S_x= \sum^N_{j=1} X_j$, give atomic level occupation in $Z$ and $X$ basis, respectively.
While the first term is the internal energy of the bosonic mode, the second is the internal energy of the atomic ensemble where $\Omega$ is the energy difference between the two atomic levels.

The third term describes the fermion-boson interaction that creates/annihilates one boson while causing a transition in atomic energy level.
The coupling between the atoms and the bosonic field is characterised by the strength $g$.
Note that the three real parameters can be combined into two dimensionless parameters, such as $\Omega/\omega$ and $g/\omega$.
The Dicke Hamiltonian commutes with the total spin operator $S^2:= S^2_x + S^2_y + S^2_z$ indicating conservation of the total spin quantum number.

After choosing
\begin{align}
H_0= \omega_0 b^\dag b \otimes \mathds{1} + g (b + b^\dag) \otimes S_x
\end{align}
we introduce the transformed operators
\begin{align}\label{eq:DickeTransformedOperators}
\tb:= b \otimes \mathds{1} + \frac{g}{\omega_0} \mathds{1} \otimes S_x.
\end{align}
As a result,
\begin{align}\label{eq:H0DickeDiag}
H_0 &= \omega_0 \tb^\dag \tb - \frac{g^2}{\omega_0} S^2_x,
\end{align}
where $n_i$ is the spin of $i$th atomic mode in the $x$-direction.
Remark that $H_0$ expressed as in Eq.~\eqref{eq:H0DickeDiag} is diagonal hence exponentially fast-forwardable.
The remaining part is the unitary that realises the transformation given in Eq.~\eqref{eq:DickeTransformedOperators}.
This unitary follows from the basic displacement operator given in Eq.~\eqref{eq:DisplacementOperator2Q}.
Due to the fact that actions of operators on the bosonic and atomic dofs commute, in second and first-quantised form, respectively,
\begin{subequations}
\begin{align}\label{eq:PolaronTransformDicke}
\mD &= \exp\left[\frac{g}{\omega_0} (b^\dag - b) \otimes S_x  \right]\\
&= \exp\left[-\ii \frac{g \sqrt{2}}{\omega_0} P \otimes S_x \right]
\end{align}
\end{subequations}
leads to the transformation given in Eq.~\eqref{eq:DickeTransformedOperators} behaving as $\tb= \mD^\dag b \mD$.
In the first-quantised form $\mD$ can be implemented with gate complexity of $\mathcal{O}(N + \plog(\Lambda \omega/(g \epsilon),N \omega/(g \epsilon)))$.
Note that this is due to the fact that $S_x$ and $P$ are both exponentially fast-forwardable.
Furthermore, recent results of Ref.~\cite{jain2025efficientquantumhermitetransform} imply that $e^{-i H_0 s}= e^{-i \omega (X^2 + P^2) s/2} e^{is (\omega/(2g))^2 \sum_{i} n_i \ketbra{n_i}{n_i}}$ can be implemented with gate complexity $N + \plog(\Lambda \omega/\epsilon, N (\omega/g)^2 |s|/\epsilon)$ hence it is also exponentially fast-forwardable.
The efficient quantum Hermite transform~\cite{jain2025efficientquantumhermitetransform} implies that, while expected to be costlier than the first-quantised form for many instances, the simulation in second-quantised form has similar asymptotic costs.
Due to the bosonic interaction connecting all spin sites we cannot apply HHKL to improve the linear scaling with the number of sites. 
Hence, without further modification to the algorithm or the quantum circuit, the cost of simulating the Dicke model for a given time $t$ is asymptotically $\mcalO(N^2 \Omega|t|)$ up to polylogarithmic factors in $N, \omega, g, 1/\epsilon$, which eliminates the polynomial dependency on the bosonic frequency $\omega$, cutoff $\Lambda$, and the coupling $g$.

\subsection{Hubbard-Fr\"{o}hlich model}\label{subsec:Frohlich}

The Fr\"{o}hlich model, introduced to capture electron-phonon interactions in Ref.~\cite{frohlich1954electrons}, is a generalisation of the Hubbard-Holstein model, where the fermion-boson interaction is not anymore only onsite, but can be long-range.
This is a natural Hamiltonian that has been devised to capture \emph{large} polarons. 
The long-range interaction is given in a functional form $f_{i,j}$ where $i$ and $j$ denote the positions of the boson and the electron, respectively.
In contrast, the Hubbard-Holstein Hamiltonian models the \emph{small} polaron hence a constant and onsite interaction, $g_{i,j}= \delta_{i,j} g$, is assumed.
The Hamiltonian terms, which are similar to those of Hubbard-Holstein but a generalisation, differ in $H_b$ and $H_{fb}$ and are given as
\begin{align}
H_b &= \sum_{i,\gamma} \omega_{i, \gamma} b^\dag_{i,\gamma} b_{i,\gamma}, \\
H_{fb} &= \sum_{i,j, \gamma} f_{i,j,\gamma}(b^\dag_{i,\gamma} + b_{i,\gamma}) (n_j - 1),
\end{align}
where $\omega_{i,\gamma}$ is the fundamental oscillator frequency of modes labeled by $\gamma$ at site $i$, and the long-range interaction between the bosonic mode $\gamma$ on site $i$ with the fermion on site $j$ is specified by an efficiently computable function $f$. 
An example form of this long-range interaction is derived in Ref.~\cite{alexandrov1999mobile} as
\begin{align}
f_{i,j}= f(|i-j|)= \frac{\kappa}{(|i-j|^3 +1)^{3/2}},
\end{align}
for some coupling $\kappa$ where there is only single bosonic mode, hence no subindex $\gamma$ is used.
Similar to the Hubbard-Holstein model, the transformation that puts $H_b + H_{fb}$ into diagonal form can be seen as a product of controlled displacement operators.
To be more precise, for a given bosonic mode $\gamma$ on site $i$, the unitary is expected to perform the following transformation
\begin{align}
\tb_{i,\gamma}= b_{i,\gamma} + \alpha_{i, \gamma} \mathds{1}
\end{align}
where
\begin{align}\label{eq:AlphaHubbardFrohlich}
\alpha_{i,\gamma} = \sum_{j} f_{i,j,\gamma}(n_j - 1)/\omega_{i,\gamma}.
\end{align}
Note that each $\alpha_{i,\gamma} \in \mathbb{R}$ depends on the fermion dofs over many sites in a diagonal manner so does not couple to any other bosonic site $j \neq i$ or $\gamma' \neq \gamma$.
Hence, in the second quantisation representation,
\begin{equation}
\mD = \prod_{i,\gamma} D(\alpha_{i, \gamma}),
\end{equation}
where
\begin{subequations}
\begin{align}
D(\alpha_{i, \gamma}) &= \exp(\alpha_{i,\gamma} (b^\dag_{i,\gamma} -  b_{i,\gamma}))\\
&= \exp(-\ii \sqrt{2}\alpha_{i, \gamma} P_{i, \gamma})
\end{align}
\end{subequations}
achieves this transformation.
The sum can then be written explicitly in terms of the polaron transform and the diagonal Hamiltonians as:
\begin{widetext}
\begin{multline}
H_b + H_{fb} =  \underbrace{\mathcal{D}^\dagger  \left(\sum_{i, \gamma} \omega_{i,\gamma} b^{\dag}_{i, \gamma} b_{i, \gamma} \otimes \mathds{1} \right) \mathcal{D}}_{H'_{b}} + \underbrace{\left(- \sum_{i, \gamma} \omega_{i, \gamma} \sum_{n_1, \ldots, n_N} |\alpha_{i, \gamma}|^2 \mathds{1} \otimes \ketbra{n_1}_1 \otimes \ldots \otimes \ketbra{n_N}_N \right)}_{H'^{\text{diag}}_{f}}.
\end{multline}
\end{widetext}
Hence the fermion-boson interaction term combined with the bosonic term is transformed by the polaron transform into a bosonic Hamiltonian, $H_{b}$, and a diagonal purely fermionic term $H'^{\text{diag}}_{f}$.
The difference to the Hubbard-Holstein case is that $\alpha_{i,\gamma}$ now depends on all sites rather than only on site $i$, yet it is an efficiently computable function whose explicit form is given in Eq.~\eqref{eq:AlphaHubbardFrohlich}.
Hence, each $\mD$ can be implemented with gate complexity $\mcalO((N_f + N_b)(\log(\Lambda \alpha /\epsilon')))$ where $\alpha:= \sum_{i, \gamma} \alpha_{i, \gamma}$ and $N_f$ and $N_b$ denote the number of sites and and number of different bosonic modes per site respectively. 
The diagonal Hamiltonian evolution with $H'^{\diag}_f$ can be implemented with gate complexity $\mcalO(\log(\max_{i} \mathcal{A}_i|s|/\epsilon'))$, where $\mathcal{A}_i:= \sum_{\gamma} \omega_{i,\gamma} |\alpha_{i, \gamma}|^2$.
This gives an overall simulation algorithm that is again polylogarithmic in the bosonic truncation parameter.
The full complexity statement will depend on the form of the coupling $f$ and $\mathcal{A}_i$ but will not affect the scaling with $\Lambda$.

\subsection{Su-Schrieffer-Heeger (SSH) model}\label{subsec:SSH}

The SSH model has been proposed in Ref.~\cite{su1980soliton} for explaining properties, in particular electrical conductivity, of doped polyacetylene chains, and reveals a mechanism that relies on topological solitons.
The Hamiltonian consists of electronic hopping terms linearly coupled to the displacement of the bosonic terms in addition to pure electronic hopping terms and a bosonic energy term.
More precisely,
\begin{align}\label{eq:HamiltonianSSH}
\nonumber H^{\text{SSH}}= &-t\sum_{i, \sigma} (c^\dag_{i+1, \sigma} c_{i,\sigma} + c^\dag_{i, \sigma} c_{i+1,\sigma})\\
&+ g \sum_{i, \sigma} (b^\dag_i + b_i)(c^\dag_{i+1, \sigma} c_{i,\sigma} + c^\dag_{i, \sigma} c_{i+1,\sigma})\\
\nonumber &+ \omega_0 \sum_{i} b^\dag_i b_i.
\end{align}

The interaction term between the boson and the fermion degrees of freedom are different to those that are in the Holstein or Fr\"{o}hlich models.
In particular, when there is only a single bosonic mode, the interaction term together with the diagonal bosonic term can be diagonalised, which is a combination of a multimode squeeze operator on the fermion degree of freedom, followed by a  transformation that appears in the Holstein model.
More precisely, the following terms
\begin{align}
H^{\text{SSH}}_{0}= \omega_0 b^\dag b + g (b^\dag + b) \sum_{i,\sigma}(c^\dag_{i+1, \sigma} c_{i,\sigma} + c^\dag_{i, \sigma} c_{i+1,\sigma})
\end{align}
are first transformed with $\mS$ where
\begin{align}
\mS^\dag c_{i, \sigma} \mS=  \frac{1}{\sqrt{N}} \sum_{q} \mathrm{e}^{\mathrm{i}qi/N} \tc_{q,\sigma}
\end{align}
which leads to
\begin{align}
\omega_0 b^\dag b + 2g (b^\dag + b) \sum_{q,\sigma} \cos(q/N) \tn_{q,\sigma},
\end{align}
where $\tn_{q,\sigma}= \tc^\dag_{q,\sigma} \tc_{q,\sigma}$ and $\tn_{q,\sigma}$ is the fermion occupation of the mode $q$ with spin $\sigma$. 
Then we can apply $\mD$ which transforms the bosonic modes controlled on the total number of fermionic modes in the new basis.
Namely,
\begin{align}
\mD^\dag b \mD = b + \frac{2g}{\omega_0}\sum_{q, \sigma} \cos(q/N) \tn_{q,\sigma} := \tb.
\end{align}
The operators $\mS$ and $\mD$ are unitary multi-mode fermionic rotation and generalised bosonic displacement operators, respectively.
Notice that $\mS$ is nothing but the fermionic fast Fourier transform (FFFT) implemented in Ref.~\cite{ferris2014fourier, kivlichan2020improved},
and $\mD$ can be implemented as a displacement operator controlled on occupation of the fermionic modes $\{q,\sigma\}$.

This method, however, fails to generalise to the original SSH Hamiltonian given as in Eq.~\eqref{eq:HamiltonianSSH} which captures a multi-mode boson, where each mode $x$ couples to the fermionic hopping between sites $i$ and $i+1$.
In fact, one can see that redefining the bosonic operator as follows: \begin{align}\label{eq:SSHPotentialTransform1}
\tb_{i,\sigma} = b_i + \frac{g}{\omega_0}(c^\dag_{i+1, \sigma}c_{i,\sigma} + c^\dag_{i, \sigma}c_{i+1,\sigma})
\end{align}
is the desired operation, that protects the bosonic algebra at each site $i$, so that fast-forwarding of the resulting diagonal operator is possible. 
However, we could not find a simple/efficient unitary transformation that achieves this for every bosonic mode $b_i$ at the same time.
This can be seen from the familiar choice of the unitary that achieves this transformation on site/mode $i$, which is a generalised displacement operator $D_i \propto \exp(b_i^\dag (c^\dag_{i+1, \sigma}c_{i,\sigma} + c^\dag_{i, \sigma}c_{i+1,\sigma}) - \textrm{h.c.})$.
Importantly, $[D_i, D_{i+1}] \neq 0$, hence simultaneous transformation of all the modes at each site $i$ with the use of a sequence of these particular displacement operators is not possible.
There is also an issue to use the polaron transform for the Tavis-Cummings model, where the interaction is
$(\omega_0/g)( b^\dag \otimes S_- - b \otimes S_+)$.
Choosing $\mD= \exp\left[\frac{g}{\omega_0} (b^\dag \otimes S_{-} - b \otimes S_{+}) \right]$ does not imply the following displacement
\begin{align}
\tb= b \otimes \mathds{1} + \frac{g}{\omega} \mathds{1} \otimes S_{-}.
\end{align}
However, particularly for the Tavis-Cummings model, the full Hamiltonian can be exponentially fast-forwardable due to the block-diagonal form provided by the generalised number operator $\mcalN= b^\dag b + S_z$.
The low dimension of each of the block also implies that one can exactly solve or classically simulate this model for a polynomially (e.g., in $\Lambda = \poly(N)$) many blocks, e.g., see Ref.~\cite{bogoliubov1996exact} for a solution via Bethe ansatz.
However, a quantum simulation and the methods presented here can still be useful when a symmetry breaking term $h$ such that $[h, \mcalN] \neq 0$ is added.

\section{Conclusions and Outlook}\label{sec:Conclusions}

We presented a substantial improvement in the simulation cost of various fermion-boson systems, where bosons are described either in first- or second-quantised form.
Our results rely on the fast-forwarding of the polaron transform, and a recent exponential fast-forwarding of the quantum harmonic oscillator Hamiltonian.
As a result, polynomial dependencies in the bosonic cutoff $\Lambda$, bosonic frequency $\omega$ and coupling strength $g$ are improved to polylogarithmic, hence an exponential improvement in (combination of) these parameters are provided.
While presented for models in one spatial dimension, our results straightforwardly generalise to higher dimensions as long as the fermion-boson interaction type is kept the same.

The immediate open question is whether these improvements can be extended beyond Hubbard-Holstein-type interactions (which additionally includes Dicke, Hubbard-Fr\"ohlich, and a single-mode SSH model).
The technical question is whether another transformation rather than a polaron transform can perform similarly, i.e., it diagonalises $H_0$, and then whether this transformation can be realised efficiently via a quantum circuit.
More fundamentally, it would be enlightening to understand under what mathematical or physical conditions purely fermionic systems are, in terms of computational complexity, any different from systems where fermions and bosons are coupled.

More practically, while the algorithm that employs IP, PT, and HHKL together is the best option in terms of asymptotic complexity, it is entirely possible to incorporate the polaron transform to any Hamiltonian simulation method, such as to QSP or even to product formulas.
Hence, it would be interesting to study the realistic regimes of cutoffs and model parameters, and get detailed resource costs for the PT incorporated directly in QSP and product formulas for small/realistic sizes.
While the intermediate ($g \sim \omega_0$) and strong ($g \gg \omega_0$) coupling regimes are classically challenging and dictate a high truncation parameter $\Lambda$~\cite{jeckelmann1998density,jansen2020finite,coulter2025electron}, it is still unclear which algorithm combined with the polaron transform performs the best in what region in the parameter space.

\section{Acknowledgments}
We thank Sam Pallister for collaboration at early stages, Sam Pallister and Mark Steudtner for discussions and feedback on the manuscript.
Correspondence should be addressed to \href{mailto:hapel@psiquantum}{hapel-at-psiquantum.com} and \href{mailto:bsahinoglu@psiquantum.com}{bsahinoglu-at-psiquantum.com}.

\newpage
\bibliographystyle{unsrt}
\bibliography{references}

\onecolumngrid 

\newpage

\begin{appendix}

\section{Fast-forwarding and block-encoding of diagonal Hamiltonians}\label{app:diag}

\subsection{Fast-forwarding of diagonal Hamiltonians}\label{appsubsec:FFDiagonalHamiltonians}

\begin{figure}[htbp]
    \centering
    \includegraphics[scale= 0.5]{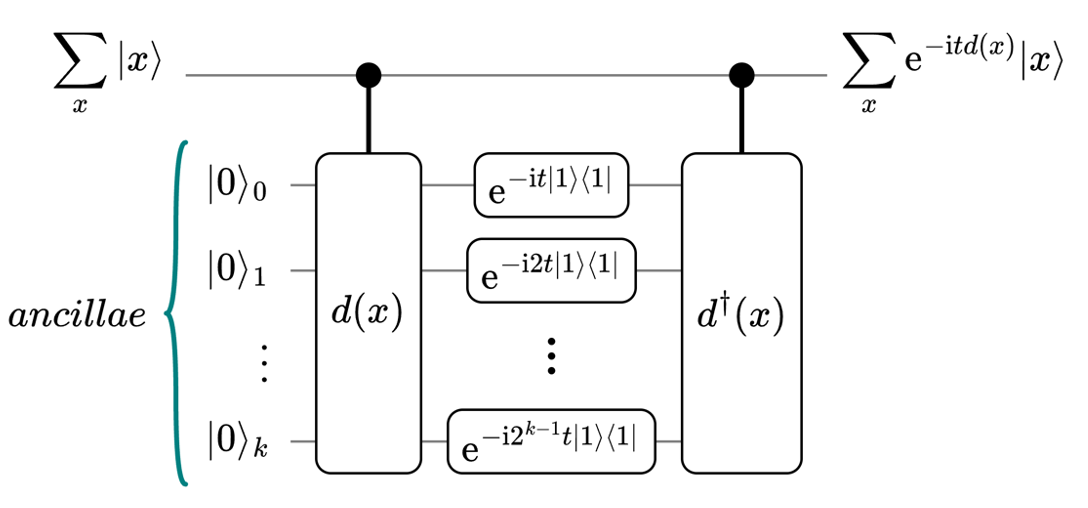}
    \caption{Circuit to simulate the evolution of a diagonal Hamiltonian $H = \sum_x d(x) \ketbra{x}{x}$ where the diagonal elements can be efficiently computed. 
    First the diagonal elements are computed in superposition onto an ancillary register with $U_{d_k}$, 
    and the $k$-qubit phase gate $R_k$ is applied to the ancilla register effectively implementing $\ket{x}\ket{d_k(x)} \rightarrow \mathrm{e}^{-\mathrm{i}t d_k(x)}\ket{x}\ket{d_k(x)}$.
    Finally the ancilla is uncomputed, disentangling it with the system register, $\mathrm{e}^{-\mathrm{i}td_k(x)}\ket{x}\ket{d_k(x)} \rightarrow \mathrm{e}^{-\mathrm{i}td_k(x)}\ket{x}\ket{0}$. 
    All together this can be implemented with gate complexity
    $\poly(k) + k \mcalO(\log_2(1/\epsilon') + \log_2 k)$, where $k = \left\lceil \log_2  \frac{1.053|t|}{\epsilon'} \right\rceil$ (assuming $\epsilon' \leq 0.1$), $U_{d_k}$ takes $\poly(k)$ and each rotation synthesis takes $\mcalO(\log_2(1/\epsilon') + \log_2 k)$ gates.}
    \label{fig:diagcirc}
\end{figure}

Diagonal Hamiltonians whose eigenvalues can be efficiently computed are a canonical case  where the evolution can be fast-forwarded, e.g. \cite{childs2004quantum}.
However, we include a description here for completeness. 
Given a Hamiltonian $H = \sum_x d(x) \ketbra{x}{x}$ the evolution operator $\mathrm{e}^{-\mathrm{i}H t}$ can be applied to a register in $\{\ket{x} \}_{x=0}^{N-1}$ basis within precision $\epsilon$ by the circuit in Fig.~\ref{fig:diagcirc} with complexity $2\mcalC(U_{d_k})) + \mcalC(R_k)$, where $U_{d_k}: \ket{x} \ket{0} \mapsto \ket{x} \ket{d(x)}$ computes $d(x)$ into a $k$-bit register and $R_k: \ket{d(x)} \mapsto e^{-\ii d(x) t} \ket{d(x)}$ is a phase gate on those $k$ qubits.
To approximate $\mathrm{e}^{-\mathrm{i}H t}$ up to error $\epsilon'$ one has to satisfy
\begin{subequations}
\begin{align}
\epsilon' &\geq |\mathrm{e}^{-\mathrm{i}t (d(x) + 2^{-k})} - \mathrm{e}^{-\mathrm{i}t d(x)}|\\
& = |\mathrm{e}^{-\mathrm{i}t 2^{-k}} - 1|\\
& \geq \tilde{\epsilon} + \frac{{\tilde{\epsilon}}^2 \mathrm{e}^{\tilde{\epsilon}}}{2}
\end{align}
\end{subequations}
where $k$ can be chosen such that $\tilde{\epsilon} \geq  |t|2^{-k}$.
For any $\epsilon' \leq 0.1$, it is sufficient to choose $\tilde{\epsilon} = 0.0951 \epsilon'$, hence $k$ can be taken as
\begin{align}
k = \left\lceil \log_2  \frac{1.053|t|}{\epsilon'} \right\rceil.
\end{align}
This implies that each of the $k$ single qubit rotations used to implement $R_k$ should be synthesised with error $\epsilon''= \epsilon'/k$.
Hence, as a result assuming that computing $d_k$ takes $\poly(k)$, the resulting quantum circuit that implements the time-evolution of a diagonal Hamiltonian takes
\begin{align}
\poly(k) + k \mcalO(\log_2(1/\epsilon') + \log_2 k).
\end{align}

\subsection{Block-encoding of diagonal Hamiltonians}\label{appsubsec:BEDiagonalHamiltonians}

A diagonal operator $A$ with efficiently computable diagonal entries, given as a function $d_A$, can be block-encoded straightforwardly, as 
\begin{align}
\BE_{A/d_{\max}} = \frac{A}{\max_x d_A(x)} \otimes \ket{\bf{0}} + \ldots \otimes \ket{\bf{0}^\perp}.
\end{align}
One way to do so is to prepare a uniform quantum state $\frac{1}{\sqrt{M}}\sum^M_{m=1} \ket{m}$, and run the inequality test $M d_A(x) \leq m d_{\max}$~\cite{lemieux2024quantum} where $d_{\max}:= \max_x {d_A(x)}$.
Then, each diagonal element of the rescaled operator is approximated with additive error at most $1/M$.
Hence,
\begin{align}
\|\BE_{A/d_{\max}} \otimes \ketbra{\bf{0}} - A \otimes \ketbra{\bf{0}}\| \leq \epsilon'
\end{align}
can be achieved by choosing $M \geq 1/\epsilon'$.
For ease of compilation, $M$ can be chosen the smallest integer power of $2$ such that $M \geq 1/\epsilon'$, i.e., $\log_2 M= \lceil \log(1/\epsilon') \rceil$.
In addition to the cost of computing the LHS of the inequality test, i.e., $d_A(x)$, the Toffoli cost of the block-encoding is then the cost of computing the RHS of the inequality test, and then realising the comparator, which is in total $\mcalO( \lceil \log_2 (1/\epsilon') \rceil \log_2(d_{\max}))$.

\section{Quantum circuits for the polaron transform in the Hubbard-Holstein model}

\subsection{Implementation of the polaron transform in first-quantised form}\label{appsubsec:DS1Q}

\begin{figure}[htbp]
    \centering
    \includegraphics[width= 0.9\linewidth]{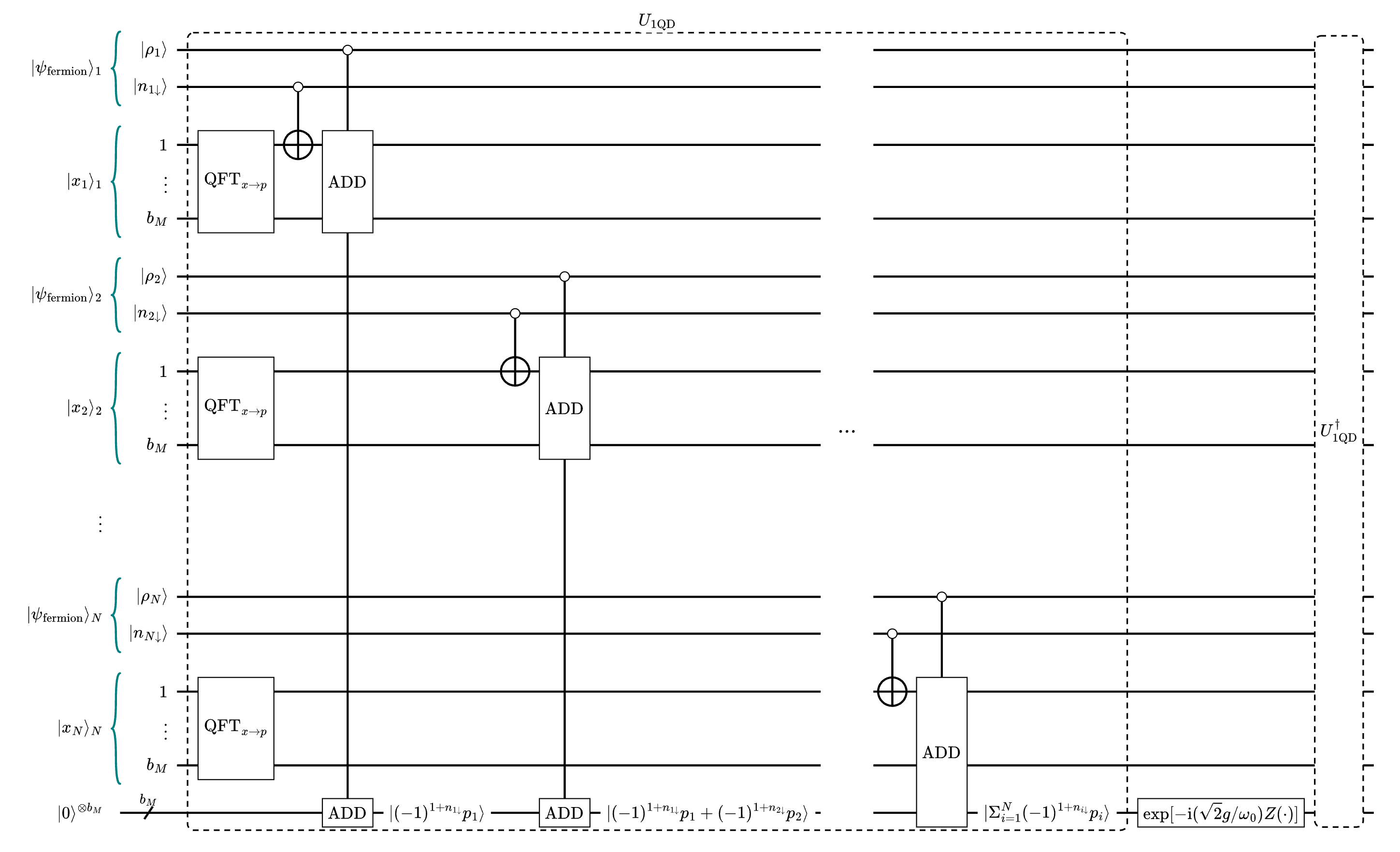}
    \caption{The circuit implementation of the polaron transform (Eqn. \ref{eqn:app dis1Q}) in first quantisation. 
    There are $i\in\{1,N\}$ sites each with a fermionic and bosonic degree of freedom. 
    The fermionic register at site $i$, $\ket{\psi_\text{fermion}}_i = \ket{\rho_i}\otimes \ket{n_{i\downarrow}}$ consists of two qubits where $\ket{\rho_i}$ holds the parity of the fermionic state and $\ket{n_{i\downarrow}}$ the down spin occupation. 
    The bosonic register at site $i$, $\ket{x_i}_i$, consists of $b_M:=\log_2(M)$ qubits and holds the position coordinate.
    $\mathrm{QFT_{x\rightarrow p}}$ denotes the centered quantum fourier transform that converts the bosonic position to momentum $\mathrm{QFT}_{x\rightarrow p}\ket{x_1}_1 = \ket{p_1}_1$- it is implemented by first shifting the computational basis with a $R_Z(\pi)$ gate on the most significant bit, followed by a normal QFT and unshifting the basis. 
    We take $M$ to be a power of $2$ so that the QFT requires $4b_M\log_2(b_M/\epsilon_{QFT})$ T gates \cite{park2024tcountoptimizationapproximatequantum}. 
    Controlled on the site's fermionic register, $(-1)^{1+ n_{i\downarrow}}p_i$ is recursively added to a $(b_N + b_M)$-ancilla register. 
    Using the implementation from \cite{Gidney_2018}, a $(b_N+b_M)$-bit adder requires $4(b_N+b_M)+\mathcal{O}(1)$ T gates, and using \cite[Lemma 7.2]{Barenco_1995} controlling such an adder on a single qubit increases this count to $16(b_N+b_M)+\mathcal{O}(1)$ T gates. 
    $(b_M + b_N)$ rotations are then performed across the ancilla to apply the correct phase before $U_{1QD}$ is uncomputed. 
    This cumulative structure for computing the phase saves a multiplicative factor of $N$.}\label{fig:disp1Q}
\end{figure}

Recall that the polaron transform for the Hubbard-Holstein model consists of controlled displacement operators and reads in first quantised form as:
\begin{equation}\label{eqn:app dis1Q}
    \mathcal{D}: = \bigotimes_i \sum_{n_x=0}^2 D_i(\alpha_{n_i})\otimes \ketbra{n_i}
\end{equation}
where
\begin{equation}
    D_i(\alpha) = \exp \left[-\mathrm{i} \alpha P_i^d \right] \qquad  \text{and again} \qquad \alpha_{n_i} = \begin{cases}
+g/\omega_0,\; &\textrm{if} \; n_i=2,\\
0 ,\; &\textrm{if} \; n_i = 1,\\
-g/\omega_0,\; &\textrm{if} \; n_i=0.
\end{cases}
\end{equation}
$P$ is simply related to $X$ via the centered QFT (Eq.~\ref{eqn:centeredQFT}) and hence can efficiently be implemented.
Figure~\ref{fig:disp1Q} gives an implementation with
\begin{equation}\label{eqn: cost 1Q circuit}
    4Nb_M\log_2(b_M/\epsilon_\text{QFT})+ 32 N(b_N+b_M) + (b_N + b_M)(0.53 \log_2(1/\epsilon_\text{rot})+4.86) \quad \text{T gates},
\end{equation}
and $(b_N + b_M)$ temporary ancilla qubits, where $b_N= \lceil \log_2 N \rceil$ and $b_M = \lceil \log_2 M \rceil$. 
The first term comes from $N$ parallel $\QFT$s, the second term the cost of the adders, and the last term the cost of the $(b_N + b_M)$-qubit rotation synthesis.
The total cost is then $\mcalO(N \log M \log (\log M/ \epsilon_{\QFT}) + N (\log N + \log M) + (N + M)\log(1/\epsilon_{\rot}))$.

\subsection{Implementations of the polaron transform in second-quantised form}\label{appsubsec:DS2Q}

We will present two implementations of the polaron transform in second-quantised form.
First uses QHT to implement the polaron transform in the first-quantised basis and then going back to the second-quantised basis with inverse QHT.
The second is based on expressing the operator as a controlled Hamiltonian simulation.
While the latter is substantially more complicated and costly, we still go through it because it may be of independent interest and would be useful for other models where $\mD$ is not simply a controlled displacement operator.
The former is substantially more efficient and simple, 
\begin{align}
    \mD_{2Q}= \Bigg[ \bigotimes_i \QHT^\dag_i \Bigg] \mD_{1Q} \Bigg[ \bigotimes_i \QHT_i \Bigg]
\end{align}
where $\mD_{1Q}$ is given in Section~\ref{appsubsec:DS1Q}.

For the second implementation, recall the displacement operator in second quantisation:
\begin{equation}
    \mathcal{D}: = \bigotimes_i \sum_{n_i=0}^2 D_i(\alpha_{n_i})\otimes \ketbra{n_i}
\end{equation}
where
\begin{equation}\label{eqn:valuesalpha}
    D_i(\alpha) = \exp \left[\alpha b_i^\dagger - \alpha^* b_i \right] \qquad  \text{and} \qquad \alpha_{n_i} = \begin{cases}
+g/\omega_0,\; &\textrm{if} \; n_i=2,\\
0 ,\; &\textrm{if} \; n_i = 1,\\
-g/\omega_0,\; &\textrm{if} \; n_i=0.
\end{cases}
\end{equation}

One can treat this operator as a time-evolution, i.e., $D_i(\alpha)= \e^{-\ii K_i \ttil}$ with generating operator
\begin{align}
K_i=  \frac{\ii(b_i^\dag - b_i)}{2\sqrt{\Lambda}}
\end{align}
and time
\begin{align}
\ttil= 2 \alpha \sqrt{\Lambda}.
\end{align}
As we see below, a circuit implementing $D_i(\alpha)$ up to error $\epsilon'$ can be achieved with gate complexity $\tilde{\mcalO}(\alpha \sqrt{\Lambda}+\log(1/\epsilon'))$ using Hamiltonian simulation by qubitisation \cite{low2019hamiltonian}.

\begin{figure}[htbp]
    \centering
    \includegraphics[width= 0.8\linewidth]{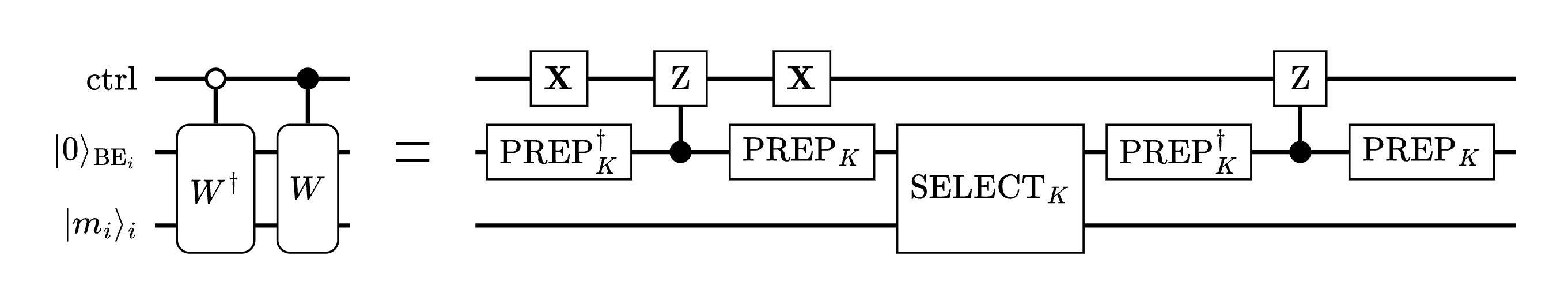}
    \caption{The controlled walk operator $c-W(K)$ in terms of prepare and select oracles.}\label{fig:walk}
\end{figure}

Note that since $\alpha$ is real, $K_i$ is independent of the fermionic register and the only dependence on the corresponding fermionic occupation is in the time through the value of $\alpha$.
Ref.~\cite{low2019hamiltonian} gives an optimal way of implementing time evolution, by first block-encoding the iterate $W(K)$ with $k \in \lceil\log_2(\Lambda)\rceil$ additional ancilla, and then combining powers of it with quantum signal processing, by using only one more additional ancilla.
Rotations are applied on this additional qubit that depend on the time $\ttil$.
As seen in the quantum circuit in Fig.~\ref{fig:walk}, there are three cases corresponding to $n_i= 2,1,0$, respectively.
In the second case no time-evolution is applied, e.g., $\mathds{1}$.
In the first and the third case a time evolution for time $\ttil = \pm 2g \sqrt{\Lambda}/\omega_0$ is applied.
This is provided by the controlled rotations with angles $\{\theta_0, \theta_1, \ldots, \theta_d\}$ and $\{\theta_0 + \theta'_0, \theta_1 + \theta'_1, \ldots, \theta_d + \theta'_d\}$, respectively.
Therefore, at a high level, one of these controlled displacement operators at a given site calls singly-controlled-$W$ $(2d+2)$ times, and singly- and doubly-controlled rotations (each) $(d+1)$ times, where $d= 2\alpha \sqrt{\Lambda} + \log(1/\epsilon')$.
The rest studies how to apply the controlled-$W$/$W^\dag$'s.

The quantum circuit for the iterate $W(K) = (2\mathrm{PREP}_K\ketbra{0}{0}\mathrm{PREP}_K^\dagger - \mathbb{I}) \cdot \mathrm{SELECT}_K$ is given by using the linear combination of unitaries, and the controlled-$W^\dag$ can be combined with controlled-$W$ as given in Fig.~\ref{fig:walk}.
Notably, the $\PREP$ and $\SEL$ subroutines don't need to be controlled, which saves constant multiplicative factors in terms of gate cost.

Expressing the generator, $K$, as a linear combination of unitaries:
\begin{subequations}
\begin{align}
    K_i &= \frac{\ii (b_i - b_i^\dagger)}{2\sqrt{\Lambda}}\\
    &= \sum_{m=1}^{\Lambda-1}\frac{\mathrm{i}\sqrt{m}}{2\sqrt{\Lambda}} \left(\ketbra{m-1}{m} - \ketbra{m}{m-1} \right).
\end{align}
\end{subequations}
It is straightforward to see that one can block-encode this operator as a special case of the block-encoding circuit given in Ref.~\cite{lemieux2024quantum}, and in Fig.~\ref{fig:walk}.
The quantum circuit uses, $\mcalO(\log_2(1/\epsilon'))$ Hadamard gates, a controlled increment and decrement of a $\log_2 \Lambda$ size system register, and an inequality test (achieved by computation of quantities related to matrix elements and a comparator) checking the following clause:
\begin{align}
&K^2 m \leq j^2 (\Lambda - 1) \quad \textrm{and} \\
&1 \leq m \leq \Lambda,
\end{align}
where $K= 2^{\lceil \log_2(1/\epsilon'') \rceil}$ is the size of the sampling space, $j$ is the auxiliary index running through the sampling space, $m$ is the bosonic occupation, and $\Lambda$ is the bosonic cutoff.
To be more precise, $\PREP$ and $\textrm{SELECT}_K$ can be chosen as in Fig.~\ref{fig:disp2Q}.
The total gate cost of these two subroutines is $\mcalO([\log_2(1/\epsilon'')]^2 \log_2 \Lambda)$.

\begin{figure}[h!]
    \centering
    \includegraphics[width= 0.9\linewidth]{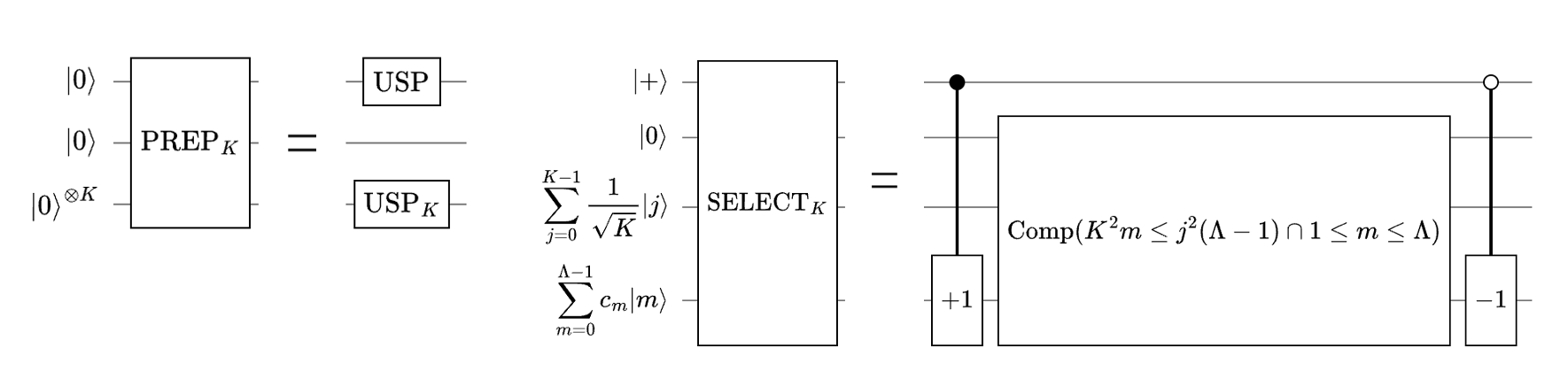}
    \caption{Implementation for $\PREP$ and $\SELECT$ for the generator of the displacement operator in second quantisation. The $\PREP$ acts on $K+2$ temporary ancilla qubits.}\label{fig:prepselect}
\end{figure}

The algorithm to implement the time evolution calls the iterate $d\in \mathcal{O}(|\tilde{t}| + \log (1/\epsilon'))$ times described in Figure~\ref{fig:disp2Q}.
This circuit is performed in parallel across all $N$ sites (for systems with geometric locality HHKL can instead be applied), 
hence the total gate complexity is: $\mcalO(N(g\sqrt{\Lambda}/\omega_0 + \log(1/\epsilon')) [\log_2(1/\epsilon'')]^2 \log_2 \Lambda )$, where $\epsilon''= \epsilon'/(2d+2)$.
\begin{figure}[htbp]
    \centering
    \includegraphics[width= 0.9\linewidth]{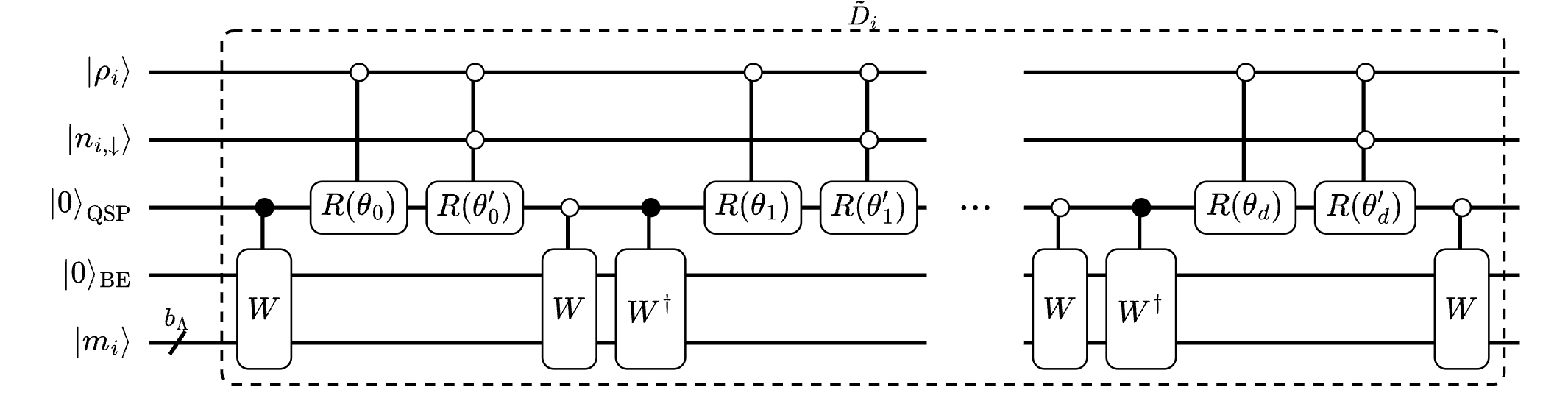}
    \caption{Circuit implementing the single site second quantisation displacement operator, $D_i(\alpha_{n_i})\otimes \ketbra{n_i}{n_i}= \e^{-\ii K_i \ttil}\otimes \ketbra{n_i}{n_i}$, via Hamiltonian simulation by qubitisation. Recall $\ket{\rho_i}$ encodes the parity of the fermionic modes and $\ket{n_{i,\downarrow}}$ the occupation of the spin down fermion at site $i$. There is a $k=\lceil \log_2(\Lambda)\rceil$ qubit ancilla for the iterate block encoding and an additional QSP ancilla to implement the exponentiation. The structure of the circuit follows the doubly efficient Hamiltonian simulation algorithm described in \cite{Berry_2024} and only uses $d \in \mathcal{O}(\tilde{t}+\log(1/\epsilon))$ calls to the iterate. The parity of the fermionic register, determines whether a non-trivial displacement operator must be applied since $D(\alpha_{n_x=1})=\mathds{1}$, hence the first rotation is controlled on this register. The second rotation is doubly controlled on both fermionic register and adjusts the angle to account for the sign of $\alpha$. The implementation of the iterate follows from Figure \ref{fig:walk}. }\label{fig:disp2Q}
\end{figure}

\section{Quantum circuits for the implementation of $\mathrm{e}^{-\mathrm{i}\tau\tH_0}$ in the Hubbard-Holstein model}
\subsection{Implementation of $\mathrm{e}^{-\mathrm{i}\tau\tH_0}$ in first quantisation via factorisation}\label{appen:H01Q}

\begin{figure}[htbp]
    \centering
    \includegraphics[width= 0.9\linewidth]{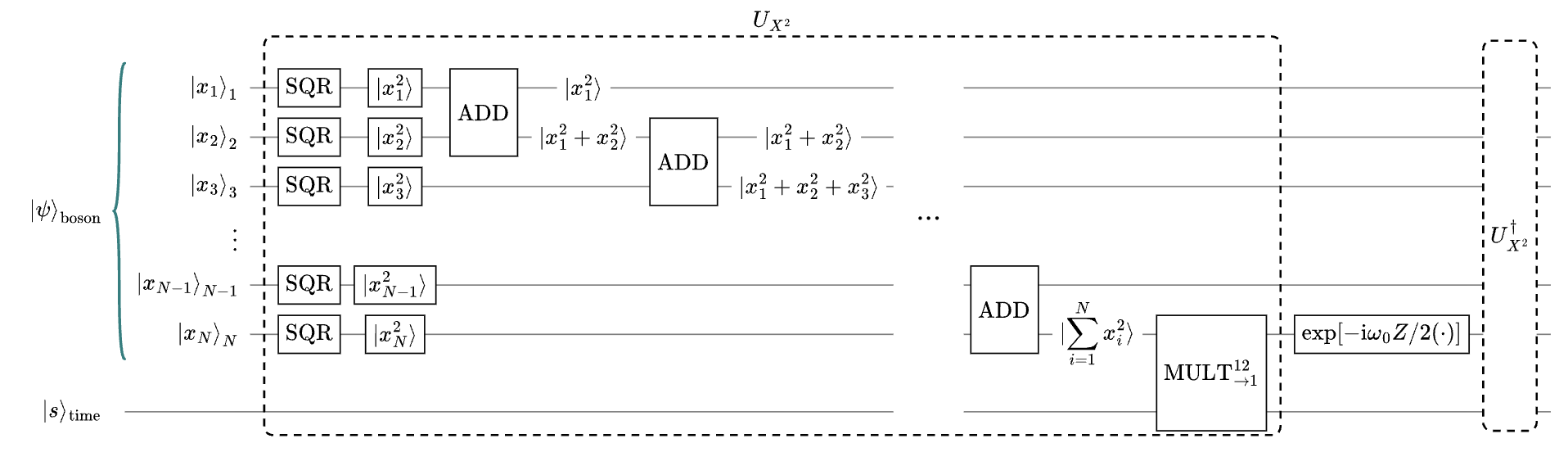}
    \caption{Circuit implementation of $\mathrm{e}^{-\mathrm{i}s X^2}$. In first quantisation each site's bosonic register consists of $b_M$ qubits and encodes the discretised position coordinate $x^d$. The first gates compute the square of the position coordinate which can be achieved with $Nb_M^2$ Toffoli. A cascade of adders collates the sum of the squared positions and costs a total of $4N(2b_M + b_N) + \mathcal{O}(1)$ T gates \cite{Gidney_2018}. $(b_N+2b_M)$ rotations are performed on the final register to apply the appropriate phase before $U_{X^2}$ is uncomputed, freeing up all ancilla for future use. }\label{fig:H01Q}
\end{figure}

For the fast-forwardable portion of the Hamiltonian in first quantisation, we will apply
\begin{align}
\e^{-\ii\tH_0 \tau}= \e^{-\ii(\sum_i X_i^2 + P_i^2) \tau} \e^{-\ii \tH^\diag_f \tau},
\end{align}
recall that $\tau \in \{l_it/rL\}_i$ is a quantum variable that depends on the value of $l_i$.
The first exponential $\e^{-\ii(\sum_i X_i^2 + P_i^2) \tau} = \bigotimes_{i=1}^N\mathrm{e}^{-\mathrm{i}\tau (X_i^2 + P_i^2)}$ will be implemented using~\cite[Thm.5]{jain2025efficientquantumhermitetransform} where they note that for any $\tau$
\begin{align}
e^{-i(P^2 + X^2)\tau} = e^{-\ii a(\tau/2) P^2} e^{-\ii b(\tau/2) X^2} e^{-2 \ii a(\tau/2) P^2} e^{-\ii b(\tau/2) X^2}e^{-ia(\tau/2) P^2}
\end{align}
where $a(t):= \tan(t/2)/2$ and $b(t):= \sin(t)/2$.
For the discretised/truncated Harmonic oscillator, Ref.~\cite{jain2025efficientquantumhermitetransform} shows that this decomposition is exponentially precise in $M= \eta \Lambda/\log \Lambda$, i.e., the approximation error decays exponentially in $M$, for large enough $M$ that grows almost linearly in $\Lambda$.
Both circuit routines, $\mathrm{e}^{-\mathrm{i}a(\tau/2) P^2}$ and $\mathrm{e}^{-\mathrm{i}b(\tau/2) X^2}$, can be implemented efficiently in $\mathcal{O}(N(\log(N)+\log(M)))$  gates.
Figure \ref{fig:H01Q} describes the implementation of $\mathrm{e}^{-\mathrm{i}sX^2}$, it follows that $\mathrm{e}^{-\mathrm{i}sP^2}$ can be built similarly with a centered QFT on all registers adding an additional $4Nb_M\log_2(b_M/\epsilon_\text{QFT})$ T gates \cite{park2024tcountoptimizationapproximatequantum}.
So $\mathrm{e}^{-\mathrm{i}s X^2}$ can be implemented with
\begin{subequations}
\begin{align}
    \text{Toffoli: }&2Nb_M^2\\
    \text{T gates: }&4N(2b_M+b_N)+(b_N + b_M)(0.53 \log_2(1/\epsilon_\text{rot})+4.86).
\end{align}
\end{subequations}

\begin{figure}[htbp]
    \centering
    \includegraphics[width= 0.9\linewidth]{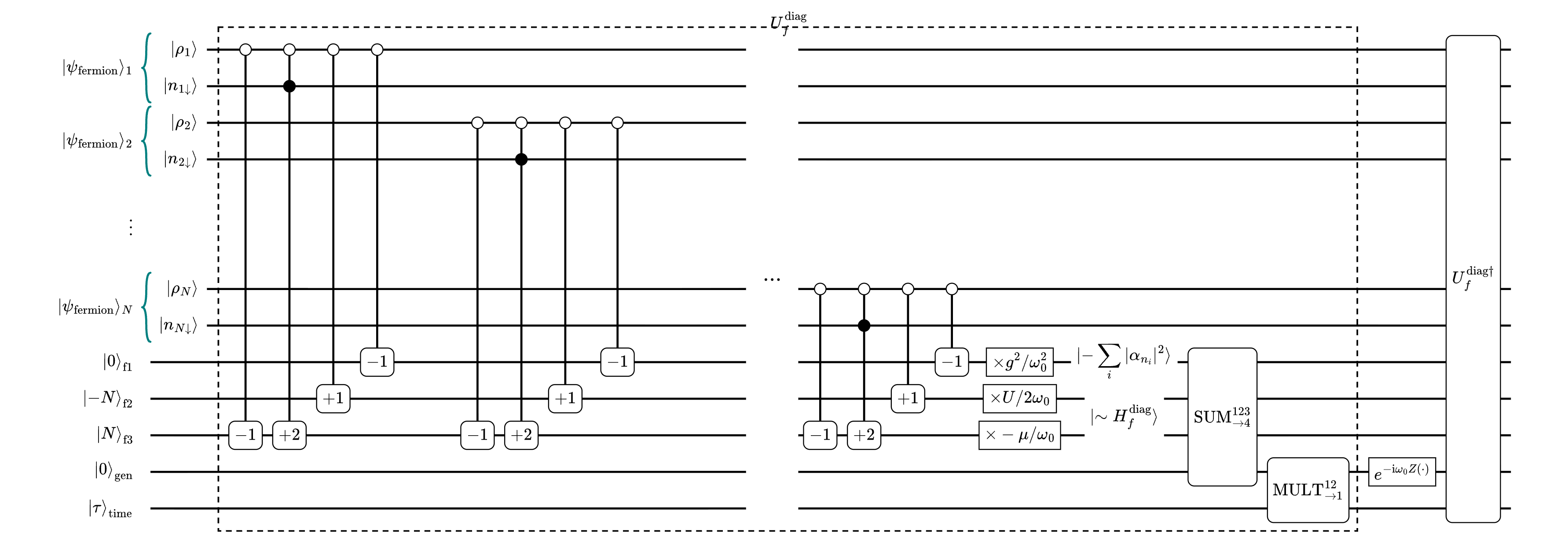}
    \caption{Circuit implementation of $\e^{-\ii \tH^\diag_f \tau}$ for first quantisation. This is a diagonal fermionic operator so can be efficiently implemented. The $-1$ denotes a decremator, and $+1$ $(+2)$ denotes incrementing by $1$ $(2)$, $\textrm{SUM}^{123}_{\rightarrow 4}$ denotes arithmetic sum of the values from $123$ and place the result in $4$. Then this summed value is multiplied by the quantum time parameter $s$ used in the Trotterisation before a phase is applied. Note that in second quantisation the $\alpha_{n_i}$ values differ from first quantisation which effects the multiplicative constants. }\label{fig:1Qfdiag}
\end{figure}

The second exponential $\e^{-\ii \tH^\diag_f \tau}$ acts only on the fermionic degrees of freedom and so commutes with the Trotterised expression and can be directly applied in parallel.
It is also diagonal in the fermionic basis so an efficient circuit exists, described in Fig.~\ref{fig:1Qfdiag}, requiring $\mcalO(N \log N + \log (N (g^2/\omega_0 + U/4 + \mu) \tau /(\epsilon \omega_0)))$ gates.
The quantum circuit in Fig.~\ref{fig:1Qfdiag} sums up the fermionic diagonal terms.
For each site $i$, fermionic dof $\ket{n_{i \downarrow}}\ket{n_{i \uparrow}}$, are already mapped into $\ket{\rho_i} \ket{n_{i \downarrow}}$ where $\rho_i= (n_{i \downarrow} + n_{i \uparrow}) \mod 2$ is the parity of the electron occupation on site $i$.
The ancilla registers labeled by subscripts $f_1$, $f_2$ and $f_3$ are used to compute the first, second and the third terms in the $\tH^\diag_f$ given as
\begin{subequations}
\begin{align}
\tH^\diag_f &= \sum_i U \left(n_{i \uparrow} - \frac{1}{2}\right) \left(n_{i \downarrow} - \frac{1}{2} \right) - \mu \sum_i n_{i} - \frac{\omega_0}{2} \sum^2_{n_{i}=0} |\alpha_{n_i}|^2 \ketbra{n_i}{n_i}\\
&\equiv \sum_{i} U \left(n_{i\uparrow} n_{i \downarrow} - \frac{1}{2} n_i \right) - \mu \sum_{i}  n_i - \frac{\omega_0}{2} \sum_{i} \sum^2_{n_i = 0} |\alpha_{n_i}|^2 \ketbra{n_i},
\end{align}
\end{subequations}
where the second line excludes the constant term.

\subsubsection{Prior state of the art Trotter based scheme for $\mathrm{e}^{-\mathrm{i}\tau\tH_0}$ in first quantisation}\label{append:prior1Q}

Prior to \cite{jain2025efficientquantumhermitetransform} the state of the art for an $\mcalO(\epsilon)$-approximate circuit implementation of exponentiated quantum harmonic oscillator $\mathrm{exp}(-\mathrm{i}((X)^2 + (P)^2)s)$  still used only $\mcalO(\exp \tilde{\gamma}\sqrt{\log(cM/\epsilon)})$ two-qubit gates for some constant $\tilde{\gamma}$.
A Trotter-Suzuki \cite{suzuki} scheme is used to approximate the full evolution by a product consisting of terms of the kind $\mathrm{e}^{-\mathrm{i}(X)^2 s}$ and $\mathrm{e}^{-\mathrm{i}(P)^2 s}$, both of which are diagonal and thus efficient.
Standard Trotter-Suzuki argument would require $\mathcal{O}(M)$ terms in the product leading to superlinear in $M$ gate complexity. 
The key of \cite{somma2016quantum} was to note that the effective norm of the nested commutators has a tighter bound than just the product of the norms since $X$ and $P$ form an approximate basis for a Lie algebra of dimension $3$ \cite[Appendix B]{somma2016quantum} and so fewer terms are required in the product.

The $p$'th order Trotter is described recursively as 
\begin{subequations}
\begin{align}
    U_1^d(s) &:= \mathrm{e}^{-\mathrm{i}(\frac{s}{4})X^2}\mathrm{e}^{-\mathrm{i}(\frac{s}{2})P^2}\mathrm{e}^{-\mathrm{i}(\frac{s}{4})X^2},\\
    U_{p+1}^d(s) &:= U_p^d(s_p)^2U_p^d(s-4s_p)U_p^d(s_p)^2
\end{align}
\end{subequations}
where $p\in \mathbb{Z}^+ \geq 2$ and $s_p = \frac{s}{4-4^{1/(2p+1)}}$.
\cite[Theorem 2]{somma2016quantum} shows that the evolution $U^d(\tau) := \mathrm{e}^{-\mathrm{i}H_0 \tau}$ is $\epsilon_\text{Trott}$-approximated by $\left[U_p^d(s)\right]^\kappa$, where $\tau = s\kappa$, for states with non-trivial support on the first $M'$ eigenstates, i.e. $\ket{\phi} = \sum_{m=1}^{M'}c_m \ket{\psi_m^d}$ (Eq. \ref{eqn: QHO eigenstates}). The total Hilbert space dimension is given by $M = \exp\{\mathcal{O}(\sqrt{\log(M'\abs{\tau})/\epsilon_\text{Trott}}) \} + \mathcal{O}(M')$.
The Trotter order $p \in \Theta (\sqrt{\log(M'\abs{\tau}/\epsilon_\text{Trott})}) $ and the size of Trotter step $\abs{s}\in\theta(5^{-p})$ are chosen following \cite{somma2016quantum} such that the the number of calls to $\mathrm{e}^{-\mathrm{i}sX^2}$, $\mathrm{e}^{-\mathrm{i}sP^2}$ is $\mathcal{M} \in \mathcal{O}(\exp(\gamma \sqrt{\log(M'/\epsilon)}))$ for constant $\gamma$. 
Hence the total gate complexity for simulating the $H_0$ is $O\left(N(\log(M)+\log(N))\exp(\gamma \sqrt{\log(M'/\epsilon)})\right)$ and subpolynomial in the bosonic cutoff $M$.

\subsection{Implementation of $\mathrm{e}^{-\mathrm{i}\tau\tH_0}$ in second quantisation}\label{appen:H02Q}

In the second quantised form, the time evolution reads as a product of two exponentials
\begin{align}
\mathrm{e}^{-\mathrm{i}\tau\tH_0} = \exp(- \ii \tau \omega_0 \sum_i b^\dag_i b_i) \exp( \ii \tau \sum_i \sum^2_{n_i = 0} |\alpha_{n_i}|^2 \otimes \ketbra{n_i}{n_i}).
\end{align}
Both of these exponentials are diagonal operators and can be exponentially fast-forwarded.
To be more precise, the first exponential can be implemented by first computing $\sum_i m_i$ to a register, and then apply a phase kick-back with a multiplying factor $\tau \omega_0$.
Implementing this unitary with $\epsilon'$ error takes $\mcalO(\max\{\log_2(|\tau \omega_0|\Lambda N/\epsilon'), \log(1/\epsilon')\} + N \log_2 \Lambda)$ gates.
The second exponential can be implemented similarly.
Computing the sum over $|\alpha_{n_i}|^2$ can be done via $\mcalO(N)$ controlled additions, which is then multiplied with $\tau$ and a phase-kickback is applied.
Implementing this unitary with $\epsilon'$ error hence takes $\mcalO( \max\{\log_2 (|\tau g^2/(\omega^2_0 \epsilon')|), \log_2 (1/\epsilon')\} + N)$.

\section{Block-encoding of the Hubbard-Holstein Hamiltonian terms}\label{app:BlockEncodingGeneral}

\begin{figure}[htbp]
    \centering
    \includegraphics[scale= 0.5]{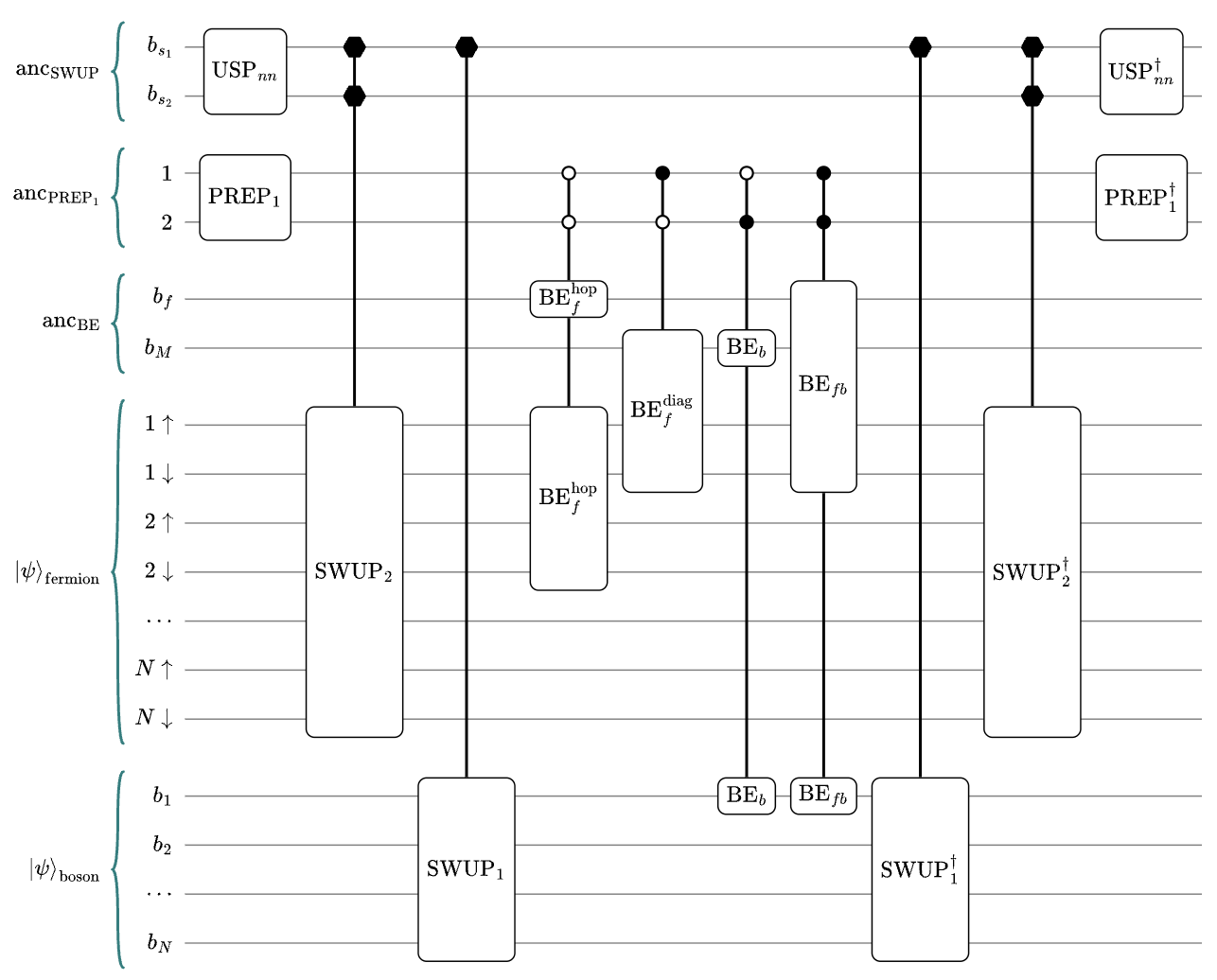}
    \caption{High level circuit construction of the block-encoding of the full Hamiltonian as a $\PREP^\dag-\SEL-\PREP$ circuit. 
    The first and second set of registers are used for creating superpositions for $\PREP$ subroutine, controlled - $\SWUP$s are part of the full $\PREP$ subroutine.
    Controlled-$\SEL$s are applied to realise the fermion, boson, fermion - boson terms if they are turned on which is determined by the $\PREP_1$ subroutine.
    The rest is uncomputation.}
    \label{fig:FullBE-HighLevel}
\end{figure}

In this section, we give various versions of block-encodings of the Hamiltonian terms in the Hubbard-Holstein model, both in first and second quantisation.
Remark that, parts of these circuits can be discarded if the method of implementation is not via block-encoding the full Hamiltonian.
In fact, we advocate in the main text that, for asymptotic efficiency at least, only the fermionic hopping term $H^{\hop}_f$ should be block-encoded and the rest is fast-forwarded and included in the Hamiltonian simulation via the interaction picture.
Nevertheless, in some parameter regimes, it could be more efficient to shift parts of the Hamiltonian to the interaction term $V$ to be block-encoded, rather than including them in $H_0$.
In that case, this appendix gives full details of particular block-encodings of each term and how they could be combined via an LCU.
The high level circuit using $\SWUP$s is given in Fig.~\ref{fig:FullBE-HighLevel}, Ref.~\cite{babbush2018encoding} gives another circuit which does not use SWUPs to exploit translation invariance, but instead uses an indexing scheme to apply the Hamiltonian as an LCU.

We give the detailed block-encoding of $H^{\hop}_f$, $H_f$, and the full Hamiltonian $H$.
We treat the fermions always in the second-quantisation, while the bosons are treated both in second- and first-quantised form.
Parts of the high-level circuit are common to all block-encodings given below.
Specifically, $\USP_{nn}$, $\SWUP_2$, and their uncomputations.
The complexity of $\PREP_1$ increases as more terms are block-encoded, and $\SWUP_1$ is used only when boson and/or boson-fermion terms are block-encoded.
$\SEL_f$, $\SEL_b$, and $\SEL_{fb}$ refer to the select operations used for block-encoding the fermion, boson, and fermion-boson terms, respectively, and are used only when these terms are block-encoded.

The unitaries $\USP_{N \choose 2}$ and $\SWUP_2$ are defined as:
\begin{align}
&\USP_{nn} \ket{0} \ket{0}:= \frac{1}{\mcalN_{nn}} \sum^{N}_{i=1} \sum^{N}_{j=1: \langle i, j \rangle} \ket{i} \ket{j},\\
&\SWUP_2 \ket{i} \ket{j} \ket{a_{1 \uparrow}, a_{1 \downarrow}, a_{2 \uparrow}, a_{2 \downarrow}\ldots, a_{i \uparrow}, a_{i \downarrow},\ldots, a_{j \uparrow}, a_{j \downarrow},\ldots} := \ket{i} \ket{j} \ket{a_{i \uparrow}, a_{i \downarrow}, a_{j \uparrow}, a_{j \downarrow}, \ldots, a_{1 \uparrow}, a_{1 \downarrow}, \ldots, a_{2 \uparrow}, a_{2 \downarrow},\ldots },
\end{align}
where $\USP_{nn}$ creates an equal weight superposition of all possible nearest-neighbouring doublets on the underlying lattice. 
Controlled on the resulting registers' values, $\SWUP_2$ swaps the $i,j$-th fermion registers to the first and second fermion registers ($\uparrow$ and $\downarrow$'s move altogether).

We implement these two subroutines as follows.

\subsubsection{An implementation of $\USP_{nn}$ and its resource cost}
\label{subsubsec:USPCost}

For $\USP_{nn}$, the first register $b_{s_1}$ is of size $b_N= \lceil \log_2 N \rceil$.
For one-dimensional nearest-neighbour interaction term we do not need the second register denoted by $b_{s_2}$.
This register is initiated in the $\ket{0}$ state. 
The following state is prepared: $\frac{1}{\sqrt{2^{b_N}}} \sum^{2^{b_N}-1}_{i=0} \ket{i}$ costing $b_N$ Hadamard gates.
We then coherently sample $\frac{1}{\sqrt{2N}} \sum^{N}_{i=1} \ket{i}$ using the quantum circuit in Ref.~\cite{lemieux2024quantum}.
This proceeds by first flagging the cases $2^{b_N}> i> N$, which is achieved by a $b_N$-bit comparator that can be implemented with $b_N$ Toffolis, and an additional temporary $(b_N+1)$ qubits.
Note that, in the worst case, one needs one round of amplitude amplification, because $\sum_{i} |g_{i}|^2/\sum_{i} |f_{i}|^2= 2^{b_N}/N< 4$. 
This implies that we need only two reflections controlled on the flag qubits, e.g., $2$ CCZ gates (CCZ = Toffoli), and a single rotation of $b_r$-bit precision.
All in all, the cost of this subroutine is $3b_N + b_r + 2$ Toffolis, and $b_N + 1$ additional ancilla, together with additional Clifford gates.

This is the construction for one dimensional lattice with nearest-neighbor kinetic term interaction.
This can be easily generalised to other interactions such as next-nearest-neighbour (nnn), and/or higher dimensions.

\subsubsection{An implementation of $\SWUP_2$ and its resource cost}
\label{subsubsec:SWUP2Cost}

$\SWUP_2$ can be similarly simplified for the one-dimensional nearest-neighbour kinetic term.
It is implemented as a multiplexed controlled-SWAP.
In general, controlled on the $\anc_{\SWUP}$ registers $b_{s_1}$ and $b_{s_2}$, it acts on the fermion registers labeled by $1\uparrow, 1\downarrow, \ldots, N\uparrow, N\downarrow$.
Controlled on the index $i$, the qubits in $i\uparrow$ and $i\downarrow$ registers are swapped with the qubits in $1\uparrow$ and $1\downarrow$ registers.
Similarly, controlled on the index $j$, the qubits in $j\uparrow$ and $j\downarrow$ registers are swapped with the qubits in $2\uparrow$ and $2\downarrow$ registers.
This is done consecutively, and costs, at most, $4N-1$ controlled-SWAPs, where each controlled-SWAP is controlled by $b_N$ qubits.
One controlled-SWAP can be implemented with one Toffoli and two CNOTs.
The control logic costs $b_N$-controlled-CNOTs.
Hence, this subroutine costs at most $(4N-1)\mathcal{O}(b_N)$ Toffolis.

Note that this is the cost of $\SWUP_2$ in the most general case, and without applying the \emph{unary iteration technique} introduced in Ref.~\cite{babbush2018encoding}.
In our case, the lattice has constant degree, specifically, it's a one-dimensional lattice with nearest-neighbour interaction.
Hence, we know that whenever we swup the $i$th register with the first register, we also and only swup the $j=i+1$st register with the second register.
Hence for the control logic that selects turns on an ancilla when the $b_{s_1}$ register is in state $\ket{i}$, we apply $4$ controlled-SWAPs, that swaps the registers $1\uparrow$, $1\downarrow$, $2\uparrow$, $2\downarrow$ with registers $i\uparrow$, $i\downarrow$, $(i+1)\uparrow$, $(i+1)\downarrow$.
As investigated in~\cite[see Sec.~IIIA]{babbush2018encoding}, the control logic costs only $N-1$ Toffolis, at the expense of introducing $b_N$ temporary ancilla qubits.
Hence the total cost of $\SWUP_2$ for this special case is $5N-1$ Toffolis.

\subsection{Block-encoding of $H^{\hop}_f$}\label{app:BEHfhop}

In this section, we give implementations and costs of $\PREP^{\hop}_f$ and $\SEL_f$, to block-encode only the term $H^{\hop}_f$ summarised in Table~\ref{tab:ResourceCostBEHhopf}.
For the purpose of block-encoding only $H^{\hop}_f$, a register $b_f= 3$ qubits is initiated in the $\ket{{\bf 0 }} =\ket{0}\ket{0} \ket{0}$ state.
This is because, for each spin $\sigma$, the hopping term $c^\dagger_{x,\sigma} c_{x+1,\sigma} + c^\dagger_{x+1,\sigma} c_{x,\sigma}$ is expressed as a sum of $4$ Pauli-products, namely,
\begin{subequations}
\begin{align}
c^\dagger_{i,\sigma} c_{i+1,\sigma} + c^\dagger_{i+1,\sigma} c_{i,\sigma} &= \frac{1}{4}[(X_{i,\sigma} - iY_{i,\sigma})\otimes Z_{i+1, \sigma}(X_{i+1, \sigma} + iY_{i+1, \sigma}) + (X_{i, \sigma} - iY_{i, \sigma})Z_{i, \sigma} \otimes (X_{i+1, \sigma} + iY_{i+1, \sigma})]\\
&= \frac{1}{4}[(X_{i,\sigma} - iY_{i,\sigma})\otimes (i Y_{i+1,\sigma} + X_{i+1, \sigma}) + (- iY_{i,\sigma} + X_{i,\sigma}) \otimes (X_{i+1, \sigma} + iY_{i+1, \sigma})]\\
&= \frac{1}{2} (X_{i,\sigma} - iY_{i,\sigma}) \otimes (X_{i+1, \sigma} + iY_{i+1, \sigma}) .
\end{align}
\end{subequations}
Hence an equal weight linear combination of these unitaries can be achieved with $\PREP^{\hop}_f$ and $\SEL_f$ defined as
\begin{align}
\PREP^{\hop}_f \ket{0} \ket{0} \ket{0} &= \frac{1}{\sqrt{2}} \sum^2_{a=1} \ket{a} \frac{1}{\sqrt{2}} \sum^2_{b=1} \ket{b} \frac{1}{\sqrt{2}} \sum_{\sigma \in \{\uparrow, \downarrow\}} \ket{\sigma}\\
\SEL_f &=  \sum^2_{a,b=1} \sum_{\sigma \in \{\uparrow, \downarrow\}}  \ketbra{a} \otimes \ketbra{b} \otimes \ketbra{\sigma} \otimes (a X_{1, \sigma} - \bar{a} iY_{1, \sigma}) \otimes (b X_{2, \sigma} + \bar{b} iY_{2, \sigma})
\end{align}
where $\bar{a}= a \oplus 1$ is the negation of the binary value $a$. 
Then, $(\PREP^{\hop}_f)^\dagger \cdot \SEL_f \cdot \PREP^{\hop}_f$, on the the $1\uparrow$, $1\downarrow$, $2\uparrow$, $2\downarrow$ registers, acts as

\begin{align}
(\PREP^{\hop}_f)^\dagger \cdot \SEL_f \cdot \PREP^{\hop}_f = \ketbra{0} \otimes \ketbra{0} \otimes \ketbra{0} \otimes \frac{\sum_{\sigma \in \{\uparrow, \downarrow\}}c^\dagger_{1,\sigma} c_{2, \sigma} + c^\dagger_{2,\sigma} c_{1, \sigma} }{4}   +  Q
\end{align}
where $Q$ is such that $\bra{{\bf 0}}Q\ket{{\bf 0}}= 0$.
Note that $\PREP^{\hop}_f = H^{\otimes 3}$, and $\SEL_f$ consists of only $2$ CNOTs sandwiched between $4$ CZs so both subroutines cost no Toffolis.
The rescaling factor of $H^{\hop}_f$ is then found to be
\begin{align}
\alpha_{H^{\hop}_f}= 4N.
\end{align}

\begin{table}[h!]
    \centering
    \setlength{\extrarowheight}{2pt}
    \begin{tabular}{|c||c|c|c|}
    \hline 
    & Toffoli & Additional ancilla & Temporary ancilla \\
    \hline
    \hline
    $\USP_{nn}$& $3b_N + b_r + 2$ & $b_N + 1$ & $b_N+1$  \\
    \hline
    $\PREP_1$& $0$ & $3$ & $0$  \\
    \hline
    $\SWUP_2$& $5N-1$ & $0$ & $b_N$  \\
    \hline
    $\SEL_f$& $0$ & $0$  & $0$  \\
    \hline
    \end{tabular}
    \caption{Resource cost of subroutines for block-encoding $H^{\hop}_f$ for the special case of nearest-neighbor interaction in one spatial dimension.
    $N$ is the number of lattice sites, $b_N= \lceil \log_2 N \rceil$, $b_r$ is the bit precision of the rotation synthesis.}
    \label{tab:ResourceCostBEHhopf}
\end{table}

When considering the transformed hopping Hamiltonian from Eq. \ref{eq:1QtildeV}, the expression above becomes:
\begin{multline}
    c^\dagger_{i,\sigma} c_{i+1,\sigma}\mathrm{e}^{-\mathrm{i}g\sqrt{2}(P_i-P_{i+1})/\omega_0} + c^\dagger_{i+1,\sigma} c_{i,\sigma}\mathrm{e}^{\mathrm{i}g\sqrt{2}(P_i-P_{i+1})/\omega_0}  \\= \frac{1}{4}(\mathrm{e}^{-\mathrm{i}g\sqrt{2}(P_i-P_{i+1})/\omega_0} + \mathrm{e}^{\mathrm{i}g\sqrt{2}(P_i-P_{i+1})/\omega_0}) (X_{i,\sigma} - iY_{i,\sigma}) \otimes (X_{i+1, \sigma} + iY_{i+1, \sigma}) .
\end{multline}
Hence the $\SELECT$ is unchanged and the $\PREP$ has an additional phasing factor.
This phase increases the complexity of $\PREP$ requiring a temporary ancilla to apply the phase and the complexity of implementing one $\mathrm{e}^{\mathrm{i}g\sqrt{2}(P_i-P_{i+1})/\omega_0}$, introducing an additive $\mathcal{O}(\log(M))$ gates.

\subsection{Block-encoding of $H_f$}\label{app:BEHf}
In this section, we add $H^{\diag}_f$ to the block-encoding of $H_f^{\text{hop}}$ which has been given in the previous section.
This requires only a slight modification of the subroutines $\PREP_1$ and $\SEL_f$, given that we now block-encode 
\begin{align}
U(n_{i,\uparrow} - 1/2) (n_{i,\downarrow} - 1/2) - \mu n_i - \sum_{\sigma \in \{\uparrow, \downarrow\}} c^{\dagger}_{i,\sigma} c_{i+1, \sigma} + c^{\dagger}_{i+1,\sigma} c_{i, \sigma}.
\end{align}

\begin{table}[h!]
    \centering
    \setlength{\extrarowheight}{2pt}
    \begin{tabular}{|c||c|c|c|}
    \hline 
    & Toffoli & Additional ancilla & Temporary ancilla \\
    \hline
    \hline
    $\USP_{nn}$& $3b_N + b_r + 2$ & $b_N + 1$ & $b_N + 1$   \\
    \hline
    $\PREP_1$& $2b_r$ & $4+b_M$ & $b_r$  \\
    \hline
    $\SWUP_2$& $5N-1$ & $0$ & $b_N$  \\
    \hline
    $\SEL_f$& $11 b_r + 9$ & $1$  & $b_r+1$  \\
    \hline
    \end{tabular}
    \caption{Resource cost of subroutines for block-encoding $H_f$ for the special case of nearest-neighbor interaction in one spatial dimension.
    $N$ is the number of lattice sites, $b_N= \lceil \log_2 N \rceil$, $b_r= \lceil \log_2(|U/4| + |2\mu| + 1/\epsilon') \rceil$ is the bit precision of the rotation synthesis and the arithmetics.}
    \label{tab:ResourceCostBEHf}
\end{table}

There are more than one way to achieve this, and we will proceed with a particular one.
Notice that the first two terms are diagonal in the basis we operate with, namely, after controlled-SWUP, the operator to block-encode is
\begin{align}\label{eq:SELdiagToCompute}
h_{n_{1 \uparrow}, n_{1 \downarrow}}= U(n_{1,\uparrow} - 1/2) (n_{1,\downarrow} - 1/2) - \mu (n_{1, \uparrow} + n_{1, \downarrow}).
\end{align}
For given $\ket{n_{1,\uparrow}} \ket{n_{1,\downarrow}}$, we block-encode the diagonal element above directly.
Namely, the $\SEL_f$ operation computes Eq.~\eqref{eq:SELdiagToCompute} to an ancilla of $b_r= \lceil \log_2(|U/4| +|2\mu|+ 1/\epsilon')\rceil$ qubits where $\epsilon'$ is the additive accuracy of the diagonal elements that is sufficient to have desired precision for the block-encoding.
We implement this part of the block-encoding via quantum rejection sampling technique.
With the following modifications to $\PREP_1$ and $\SEL_f$, we implement the block-encoding of $H_f$:
\begin{align}
\PREP_1 \ket{0} =  \frac{1}{\sqrt{\alpha_f}}\left(\sqrt{\alpha^{\hop}_{f}} \ket{0} + \sqrt{\alpha^{\diag}_{f}} \ket{1} \right),
\end{align}
where $\alpha^{\hop}_f=4$ and $\alpha^{\diag}_f= |U/4|+ |2\mu|$, and $\alpha_f= 4 + |U/4|+ |2\mu|$.
$\SEL_f$ now consists of two $\SEL^{\hop}_f$ and $\SEL^\diag_f$, which are given as
\begin{align}
\PREP_f \ket{0}\ket{0} \ket{0} \ket{0}&=  \frac{1}{\sqrt{2}} \sum^1_{a=0} \ket{a} \frac{1}{\sqrt{2}} \sum^1_{b=0} \ket{b} \frac{1}{\sqrt{2}} \sum_{\sigma \in \{\uparrow, \downarrow\}} \ket{\sigma} \frac{1}{\sqrt{M}}\sum^M_{m=1} \ket{m}\\
\SEL^{\hop}_f &= \sum^2_{a,b=1} \sum_{\sigma \in \{\uparrow, \downarrow\}}  \ketbra{a} \otimes \ketbra{b} \otimes \ketbra{\sigma} \otimes \ketbra{0} \otimes (a X_{1, \sigma} - \bar{a} iY_{1, \sigma}) \otimes (b X_{2, \sigma} + \bar{b} iY_{2, \sigma})\\
\SEL^{\diag}_f &= \mathds{1} \otimes \mathds{1} \otimes \mathds{1} \otimes \ketbra{1} \otimes \COMP
\end{align}
where $\COMP$ acts on the system registers $1 \uparrow$, $1\downarrow$ and the last ancilla register that holds the value $m$.
It first evaluates Eq.~\eqref{eq:SELdiagToCompute} to a temporary ancilla, then flags those $\ket{m}$ such that $M \alpha^{\diag}_f \leq m |h(n_{1 \uparrow}, n_{1 \downarrow})|$ with $\ket{0}$ on an additional flag ancilla qubit.
Controlled on the sign qubit of $h(n_{1 \uparrow}, n_{1 \downarrow})$, an additional $-Z$ gate is applied on the flag.
Then, $h$ is uncomputed, and $\PREP^\dagger_1$ is applied.

The cost of $\PREP_1$ is modified to $b$ Toffolis for synthesising the arbitrary angle rotation on the $4$th ancilla qubit.
Together with the uncompute, the cost is $2b$ Toffolis.
The cost of $\SEL_f$ is the sum of the costs of $\SEL^{\hop}_f$ and $\SEL^{\diag}_f$.
$\SEL^{\hop}_f$ costs $4$ Toffolis sandwiched with CZ gates.
$\SEL^{\diag}_f$ costs $b_r+1$ Toffolis for comparator, and $4b_r + b_r + 2$ Toffolis for computing the first and the second term of $h$ (each costs $\leq 2b_r$ Toffolis), and then subtracting the second from the first terms (which costs $b_r+2$ Toffolis).
Together with the uncompute of $h$, $\SEL^{\diag}_f$ costs at most $11b_r + 5$ Toffolis.
All of these costs are summarised in Tab.~\ref{tab:ResourceCostBEHf}.
The rescaling factor of $H_f$ is then found to be
\begin{align}
\alpha_{f}= N(4 + |U| + 2|\mu|).
\end{align}

\subsection{Block-encoding of $H$ in second quantised form}\label{app:BEHH2Q}

If desired, such as using QSP for implementing $\mathrm{e}^{-\mathrm{i}Ht}$ directly, one could block-encode the full Hamiltonian $H$.
This is achieved by further modifying the quantum circuit, and is given in Fig.~\ref{fig:FullBE-HighLevel} as a high-level quantum circuit, without compilation optimisation.
$\USP_{nn}$ and $\SEL_f$ are the same as in previous sections, while $\PREP_1$ is modified to include the LCU over $H_b$ and $H_{fb}$.
We then introduce and study the cost of the new components, $\SWUP_1$, $\BE_b$, and $\BE_{fb}$. 

$\PREP_1$ acts as follows:

\begin{align}
\PREP_1 \ket{0} \ket{0} = \frac{1}{\sqrt{\alpha_H}}\left(\sqrt{\alpha^{\hop}_{f}} \ket{00} + \sqrt{\alpha^{\diag}_{f}} \ket{01} + \sqrt{\alpha_{b}} \ket{10} + \sqrt{\alpha_{fb}} \ket{11} \right),
\end{align}
where $\alpha_H = \alpha_{f} + \alpha_b + \alpha_{fb}$, and $\alpha_H= N \alpha_H$ is the total rescaling factor.

$\SWUP_1$ swaps the $j$th boson register $b_j$ with the first boson register $b_1$, controlled on the index $j$ in the $\anc_{\SWUP}$ register $b_{s_1}$.
Namely, it acts as
\begin{align}
\SWUP_1= \sum^N_{j=1} \ketbra{j}_{b_{s_1}}  \otimes \sum^{\Lambda}_{m, m'=0} \ketbra{m'}{m}_{b_1} \otimes \ketbra{m}{m'}_{b_j}.
\end{align}
This is nothing but the same type of controlled-SWAP gate applied for fermion registers via $\SWUP_2$.
Optimally, we do not repeat the same control structure  that is already constructed for $\SWUP_2$. 
$\SWUP_1$ and $\SWUP_2$ are  applied consecutively under the same control logic that is by itself constructed via the \emph{unary iteration technique} of Ref.~\cite{babbush2018encoding}.
The control logic costs $N-1$ Toffolis and $b_N$ temporary ancilla qubits.
$\SWUP_2$ costs $4N$ Toffolis, and $\SWUP_1$ costs $N \lceil \log_2 \Lambda \rceil$ Toffolis.
Hence, in total $\SWUP_1$ and $\SWUP_2$ costs $N(5 + \lceil \log_2 \Lambda \rceil) - 1 $ Toffolis at the expense of $b_N$ temporary ancilla qubits.

$H_b$ and $H_{fb}$ are block-encoded in the second-quantised form as follows.
For both, $\SWUP$ takes care of implementing LCU over different sites.
$H_b$ on each site is a diagonal operator already, and can be block-encoded using the circuit given in Appendix~\ref{appsubsec:BEDiagonalHamiltonians}, using $\mcalO{(\lceil \log_2 (1/\epsilon') \rceil \log_2 (\omega_0 \Lambda))}$ gates.
The rescaling factor is $\alpha_b= \omega_0 \Lambda$.
$H_{fb}$ acts on each site as $g (b^\dag + b) \otimes (\hat{n} - 1)$, where the bosonic and the fermionic part can be block-encoded in parallel via $\mcalO{(\lceil \log_2 (1/\epsilon') \rceil \log_2 (\omega_0 \Lambda))}$ and $\mcalO(1)$ gates, respectively.
The rescaling factor is $\alpha_{fb}= g \sqrt{2 \Lambda} $.
Hence, the rescaling factor of this block-encoding of $H_b + H_{fb}$ in the second quantised-form is
\begin{align}
\alpha^{2Q}_{H_b + H_{fb}}= N(\alpha_b + \alpha_{fb})= N \sqrt{\Lambda} (\sqrt{\Lambda}\omega_0 + g\sqrt{2}). 
\end{align}
The total cost is summarised in Table~\ref{tab:ResourceCostBEH2Q}.

\begin{table}[h!]
    \centering
    \setlength{\extrarowheight}{2pt}
    \begin{tabular}{|c||c|c|c|}
    \hline 
    & Toffoli & Additional ancilla & Temporary ancilla \\
    \hline
    \hline
    $\USP_{nn}$& $3b_N + b_r + 2$ & $b_N + 1$ & $b_N + 1$   \\
    \hline
    $\PREP_1$& $2b_r$ & $4+b_M$ & $b_r$  \\
    \hline
    $\SWUP_2$ \& $\SWUP_1$ & $N(5 + \lceil \log_2 \Lambda \rceil) - 1$ & $0$ & $b_N$  \\
    \hline
    $\SEL_f$& $11 b_r + 9$ & $1$  & $b_r+1$  \\
    \hline
    \end{tabular}
    \caption{Resource cost of subroutines for block-encoding $H$ for the special case of nearest-neighbour interaction in one spatial dimension, and when the bosons are treated in second-quantised form.
    $N$ is the number of lattice sites, $b_N= \lceil \log_2 N \rceil$, $b_r= \lceil \log_2(|U/4| + |2\mu| + 1/\epsilon') \rceil$ is the bit precision of the rotation synthesis and the arithmetics.
    $\Lambda \in \mathbb{N}^+$ is the cutoff of the bosonic degree of freedom.}
    \label{tab:ResourceCostBEH2Q}
\end{table}

\subsection{Block-encoding of $H$ in first quantised form}\label{app:BEHH1Q}

The electronic part of the Hamiltonian is the same, and first-quantised form only affects the $H_b$ and $H_{fb}$ terms.
For both $H_b$ and $H_{fb}$ SWUP takes care of implementing LCU over different sites.
Then, one needs to block encode $\omega_0(X^2 + P^2)/2$ for the bosonic term and $g \sqrt{2} X \otimes (\hat{n} - 1)$.
The former can be block-encoded as an LCU of $X^2$ and $P^2= \QFT^\dag X^2 \QFT$ in the $X$-basis.
Following Appendix~\ref{appsubsec:BEDiagonalHamiltonians}, this gives rise to a rescaling factor of $M^2 \omega_0$, and costs $\mcalO(\log M)^2$ gates due to taking square and QFT (and linear in $\log M$ due to comparator).
Similarly, $X \otimes (\hat{n} - 1)$ is a diagonal operator to be block-encoded, hence following Appendix~\ref{app:diag} gives rise to a rescaling factor of $g\sqrt{2} M$, and costs $\mcalO{\log M}$ gates, due to comparator.
Given that there is $N$ of these terms, the total rescaling factor of $H_b + H_{fb}$ in first-quantised form is
\begin{align}
\alpha^{1Q}_{H_b + H_{fb}}= N (\alpha_b + \alpha_{fb})=  NM(M\omega_0/2 + g\sqrt{2}).  
\end{align}
\end{appendix}

\end{document}